\documentclass[prd,aps,epsf,psfig,twocolumn,superscriptaddress,showpacs]
{revtex4}
\usepackage{dcolumn}
\usepackage{epsfig}
\usepackage{graphicx}
\begin{document}

\vspace{2.0cm}
\title{
{ 
Simple Behavior of Primary Cross Sections for Low Mass Particles\\ 
in $p \bar p$ Collisions at y=0 and $\sqrt s =1.8$ TeV
}}
%xxxxxxxx xxxxxxxxx xxxxxxxxx xxxxxxxxx xxxxxxxxx xxxxxxxxx xxxxxxxxx xxxxxxxxxZ
% Author list goes here as well as the abstract
\author{T. Alexopoulos}
\thanks{Current address: Department of Physics, National Technical University 
of Athens, Athens, Greece}
\affiliation{Department of Physics, University of Wisconsin, Madison,
 WI 53706, USA}
\author{E.W. Anderson}
\affiliation{Department of Physics, Iowa State University, Ames,
 IA 50011, USA}
\author{A.T. Bujak}
\affiliation{Department of Physics, Purdue University, West Lafayette,
 IN 47907, USA}
\author{D.D. Carmony}
\affiliation{Department of Physics, Purdue University, West Lafayette,
 IN 47907, USA}
\author{A.R. Erwin}
\affiliation{Department of Physics, University of Wisconsin, Madison,
 WI 53706, USA}
\author{C. Findeisen}
\affiliation{Department of Physics, University of Wisconsin, Madison,
 WI 53706, USA}
\author{K. Gulbrandsen}
\affiliation{Department of Physics, University of Wisconsin, Madison,
 WI 53706, USA}
\author{L.J. Gutay}
\affiliation{Department of Physics, Purdue University, West Lafayette,
 IN 47907, USA}
\author{A.S. Hirsch}
\affiliation{Department of Physics, Purdue University, West Lafayette,
 IN 47907, USA}
\author{C. Hojvat}
\affiliation{Fermi National Accelerator Laboratory, Batavia,
 IL 60510, USA}
\author{J.R. Jennings}
\affiliation{Department of Physics, University of Wisconsin, Madison,
 WI 53706, USA}
\author{C. Loomis}
\affiliation{Department of Physics, Duke University, Durham, NC 27706, USA}
\author{J.M. LoSecco}
\affiliation{Department of Physics, University of Notre Dame, Notre Dame,
 IN 46556, USA}
\author{K.S. Nelson}
\affiliation{Department of Physics, University of Wisconsin, Madison,
 WI 53706, USA}
\author{S.H. Oh}
\affiliation{Department of Physics, Duke University, Durham, NC 27706, USA}
\author{N.T. Porile}
\affiliation{Department of Chemistry, Purdue University, West Lafayette,
 IN 47907, USA}
\author{R.P. Scharenberg}
\affiliation{Department of Physics, Purdue University, West Lafayette,
 IN 47907, USA}
\author{F. Turkot}
\affiliation{Fermi National Accelerator Laboratory, Batavia, IL 60510, USA}
\author{W.D. Walker}
\affiliation{Department of Physics, Duke University, Durham, NC 27706, USA}
\author{C.H. Wang}
\affiliation{Department of Physics, Duke University, Durham, NC 27706, USA}
\author{J. Warchol}
\affiliation{Department of Physics, University of Notre Dame, Notre Dame,
 IN 46556, USA}

%xxxxxxxx xxxxxxxxx xxxxxxxxx xxxxxxxxx xxxxxxxxx xxxxxxxxx xxxxxxxxx xxxxxxxxxZ
\begin{abstract}
A set of inclusive cross sections at zero 
rapidity, $d\sigma/dy|_{y=0}$, is presented 
for $\bar p p$ interactions at center of mass energy
$\sqrt{s}=1.8$ TeV. Six elementary particle cross sections 
are corrected
for secondary contributions from decays
of higher mass resonances in order to produce a set
of primary cross sections. The primary cross
sections per spin state 
are well described by 
$d\sigma^p/dy|_{y=0} =0.721 \cdot (\pi \lambdabar_{\pi}^2) \cdot e^{-m/T}$,
where m is the particle rest mass, $T=\hbar c/r_h$, 
and $r_h=0.97$ fm. The deuterium production cross section 
is also described if $r_h$ is 
replaced by $r_A=r_h A^{1/3}$. The same exponential in m and T describes 
primary charm fractions in
$e^+e^-$ collisions at least up to the $J/\psi$ mass. There is no significant 
evidence for strangeness or
charm suppression if only primary production of light hadrons is
considered. There is evidence that the 
primary cross section for each particle
may have the same value for pp and $\bar p p$ collisions and that it may 
have nearly constant values
between $\sqrt{s}=63$ GeV and $\sqrt{s}=1800$ GeV. Fits to the final state 
transverse momenta of the particles using a gas
model favor a temperature T=132 MeV, a chemical potential $\mu=129$ MeV, 
and a transverse flow of the
gas with $\beta_f=0.27$.
\end{abstract}


%\keywords{ }  %if this is used, one must insert the class showkeys
              %in the \documentclass{showpacs,showkeys}{revtex4}
\maketitle

\section{Introduction}

  The experiment E735 at Fermilab studied 
$p \bar p$ collisions
in the Tevatron Collider at center of mass energy $\sqrt s =1.8$ TeV.
The production of particles and their $p_t$ spectra at rapidity y=0 
were recorded along with associated multiplicity and any other
accessible information. 
These results have been published in several 
articles~\cite{b1,b2,b3,b4,b5,b6,b7,b8}. 
A few of the production rates were reported as absolute inclusive
cross sections, while others were reported as rates relative to
some simultaneously observed production processes. The purpose
of this article is to assemble the various rate data
and to present them all as absolute cross sections. 
Such a presentation reveals patterns which are relevant to
specific production models such as the statistical model, 
string model, parton
collision models, or the
quark gluon plasma.  

\section{Data Analysis}

 During the data taking period the accelerator operators made
periodic measurements of accelerator parameters with 
flying wires~\cite{b37} and wallcurrent monitors~\cite{b38}
to enable a calculation of
our luminosity 
as a function of time.
A careful analysis of the luminosity data was subsequently tabulated
by Gelfand, Grosso-Pilcher, and White~\cite{b9,b10} 
and made available to experimentalists.

 The luminosity data can be used to obtain an absolute cross section
for the production of any one particle species  
from the relation

%xxxxxxxx xxxxxxxxx xxxxxxxxx xxxxxxxxx xxxxxxxxx xxxxxxxxx xxxxxxxxx xxxxxxxxxZ

\begin{equation}
N= \sigma \int^T_0 {\cal L}(t)\,dt, \label{lum}
\end{equation}

\noindent
where ${\cal L}(t)$ is the measured luminosity as a function of time, 
$\sigma$ is some
cross section of interest, and $N$ is the number of 
events which are associated with that cross section 
during the time interval $T$.
We have previously used
the luminosity tabulations together with Eq.(\ref{lum}) to calculate
and publish invariant production cross sections for photons~\cite{b5},
phi mesons~\cite{b6}, deuterium, and tritium~\cite{b7}.    
Luminosity issues are discussed in more detail later in this article.

We have repeated our previous
analysis of $\pi$, $K$, and $p$ inclusive production with some
refinements that were intended to reduce background,
minimize tracking losses, and improve
resolution in the time-of-flight mass spectra~\cite{b11}. The new
$\pi/p$ ratio remained within 1.3\% of that reported in 
previous publications, but the relative charged kaon flux increased
$\sim$1.4 standard deviations with the new analysis. Raw data were corrected for
detector and trigger efficiencies, trigger pre-scaling, trigger dead time, and
particle loss in 0.09 absorption lengths of the spectrometer 
arm before the last time-of-flight wall at 4 meters. 
Absolute cross sections $d\sigma/dy|_{y=0}$ were calculated for these
three positively charged particles.

\section{Measured Cross Sections}

   Table \ref{sig_table} gives a list of measured single particle 
cross sections as collected from various sources. 
Column 1 lists the particles studied. Antiparticle cross sections were
the same as particle cross sections within errors. However, the cross
section for antitritium was not measurable. Errors listed here are the
result of convolution of statistical and systematic errors. 
Information on the 
associated particle multiplicity and the invariant $p_t$ spectra 
$d\sigma/dy\,dp_t^2$ for all these particles
can be found in references cited in the table. 

   The cross section listed for inclusive single photon production 
is twice the inclusive cross section
of a charged pion within errors. If one assumes production of a neutral pion is
equal to production of a charged pion, this value is consistent
with the assertion in reference \cite{b5} that single photon
production originates with the two-photon decay of the neutral pion
and possibly some small admixture of eta decay.

   In calculating invariant cross sections for the
$\Lambda^0$ and $\Xi^-$, we make use of 
particle ratios in reference \cite{b8} such as 
$[(\Lambda^0 + \bar \Lambda^0)/(p + \bar p)]$ ``in the spectrometer arm''.
Although it is not explicitly stated in reference \cite{b8}, these
actually are \emph{rapidity} based ratios~\cite{b12} such as
$[dN(\Lambda^0 + \bar \Lambda^0)/dy|_{y=0}]/ [dN(p+\bar p)/dy|_{y=0}],$
where $dN/dy$ is the number of particles per unit rapidity.

\begin{table}[h]
\caption{\label{sig_table} E735 values of $d\sigma/dy$ at rapidity 
$y=0$ for various particles produced in $p \bar p$ collisions
at $\sqrt{s}=1.8$ TeV.}
\begin{ruledtabular}
\begin{tabular}{cclc}
  Particle            & $d \sigma /dy |_{y=0}$         & Reference \\
\hline
  $\gamma$            & 114.0 $\pm$ 7.8 \rm{mb}        & \cite{b5} \\
  $\pi^+$             &  56.0 $\pm$ 6.3 \rm{mb}        & this paper \\
  $K^+$               &   7.37 $\pm$ 0.79 \rm{mb}      & this paper\\
  $p$                 &   4.37 $\pm$ 0.62 \rm{mb}      & this paper \\
  $\phi$              &   0.763 $\pm$ 0.202 \rm{mb}    & \cite{b6} \\
  $\Lambda^0$         &   1.77 $\pm$ 0.41 \rm{mb}      & \cite{b8} \\
  $\Xi^-$             &   0.282 $\pm$ 0.082 \rm{mb}    & \cite{b8}\\
  $d^+$               &   2.02 $\pm$ 0.49 $\mu$b       & \cite{b7} \\
  $t^+$               &   0.83 $\pm$ 0.34 $\mu$b       & \cite{b7} \\
\end{tabular}
\end{ruledtabular}
\end{table}

 In the literature one frequently encounters experimental
particle production ratios that 
appear to be defined by the angular aperture of a detector element. For such 
cases the ratios quoted are more likely to be formed with $dN/d\eta$ rather 
than with $dN/dy$, where $d\eta$ is an interval of pseudorapidity
which depends only on production angle and not particle mass. 
Other discussions of particle production ratios make use of the
scaling variable $x=2p_{\ell}/\sqrt{s}$ to compute $dN/dx$, where $p_{\ell}$
is the longitudinal component of momentum in the center of mass system.

In order to compare 
our data with such measurements one can use the approximate conversion
factors in Table \ref{eta_table} to convert our values of $dN/dy$ to
$dN/d\eta$ or to $dN/dx$. In this table we present conversion factors
$c_{\eta}$ and $c_x$ that can be used to multiply our values of $dN/dy$
for an approximate conversion to $\eta$ or $x$ variables at y=0.
\begin{equation}
 (dN/dy)_{y=0}=(1/c_{\eta})(dN/d\eta)_{\eta=0}=(1/c_x)(dN/dx)_{x=0}
 \label{conversion}
\end{equation}
where \newline
$c_{\eta}=p_t/\sqrt{p_t^2+m^2}$ and
$c_x=(\sqrt{s}/2)/\sqrt{p_t^2+m^2}$.

The values of $c_{\eta}$ and $c_x$ are approximate to the
extent that we have assumed $p_{\ell}=0$ and 
used observed average values of $p_t$ for each 
particle mass to estimate in Table \ref{eta_table} 
the conversion of $dy$ to equivalent $d\eta$ 
and $dx$ intervals in our experiment.

\begin{table}[h]
\caption{\label{eta_table} Columns three and four list factors by which
one can multiply $dN/dy$ in this experiment to obtain 
approximate values of $dN/d\eta$  and $dN/dx$ for the particle masses 
listed in column one. The average values of particle $p_t$ listed
in column two were used to evaluate $c_{\eta}$ and $c_x$.}
\begin{ruledtabular}
\begin{tabular}{cccc}
mass (GeV/c$^2$)  &  $<\!\!p_t\!\!>$ (GeV/c)   & $c_{\eta}$& $c_x$ \\
\hline
 $\gamma(0)$      &  $0.192 \pm 0.007$ & 1.000     & 4688  \\
 $\pi(0.139)$     &  $0.433 \pm 0.020$ & 0.952     & 1978  \\
 $K(0.494)$       &  $0.590 \pm 0.020$ & 0.768     & 1170  \\
 $p(0.938)$       &  $0.716 \pm 0.030$ & 0.607     & 763   \\
 $\phi(1.019)$    &  $0.94 \pm 0.26$   & 0.678     & 649   \\ 
 $\Lambda(1.116)$ &  $0.750 \pm 0.025$ & 0.558     & 670   \\
 $\Xi(1.321)$     &  $0.90^{+0.35}_{-0.22}$ & 0.563& 563   \\
\end{tabular}
\end{ruledtabular}
\end{table}

\section{Cross Section Characterization}

\begin{figure}[h]
\psfig{figure=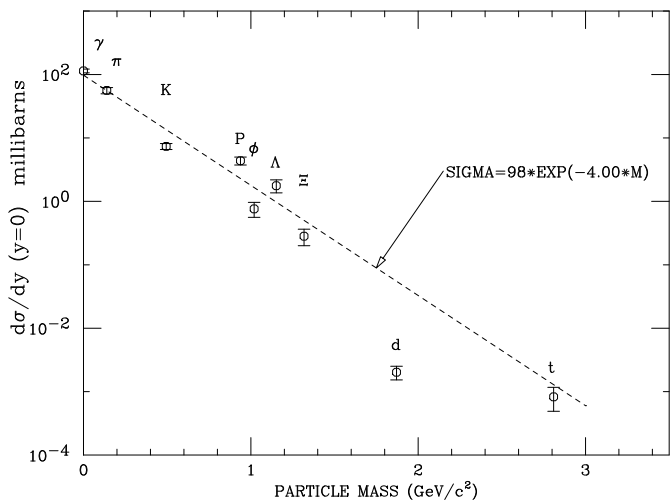,height=3.5in,angle=90}
\caption{Cross sections $d\sigma/dy|_{y=0}$ from 
Table \ref{sig_table}
plotted as a function of particle mass. The dashed
line is an arbitrary exponential falloff inserted to guide the eye.}
\label{sig_plt1}
\end{figure}

  Figure \ref{sig_plt1} is a plot of production cross sections 
versus particle mass from Table \ref{sig_table}.  
One observes what appears to be a rough
exponential decrease in cross section as the particle mass
increases. This decrease is not just some overall energy or phase space
limitation. At most about 200 charged particles are observed to 
be created in a few
of the highest multiplicity collisions when the available energy would allow
creation of around 6,000 pions, each with a kinetic energy of
a pion rest mass. Most of the non diffractive particle production is far from 
encountering a simple phase space limitation. 

  Some features of Fig.~\ref{sig_plt1} are noteworthy. The photon
cross section is not an independent data point since it is mostly 
attributable to $\pi^0$ production~\cite{b5}. Likewise electron/positron
production is negligible. One can observe electrons in the small
momentum window between 50 MeV/c, below which the field of the analysis
magnet rejects all charged particles, and 200 MeV/c, above which pions
begin having enough energy to reach the final time-of-flight counters. 
The electron mass peak is sharply defined, but it disappears completely
above 250 MeV/c underneath the low mass tail of the pion mass peak.
The observed part of the electron flux is consistent with photon 
conversions in the
2 mm thick aluminum wall of the beam pipe. Since electrons and photons
do not participate in the exponentially falling  behavior of the  
cross sections, we will omit them from most of the remaining discussions.

Although the two nuclear cross sections (d and t) are small, as the trend
for elementary particles suggests, their behavior may involve unique
nuclear effects. We will defer treatment of the nuclear cross sections 
in a quantitative way for a later section.

%xxxxxxxx xxxxxxxxx xxxxxxxxx xxxxxxxxx xxxxxxxxx xxxxxxxxx xxxxxxxxx xxxxxxxxxZ
\section{Strangeness Suppression Models}

  Most strange particle production in Fig.~\ref{sig_plt1} appears systematically
suppressed relative to non-strange production even after allowing
for some exponential decrease of cross sections with mass. 
Efforts to characterize 
this strangeness suppression have a long experimental and theoretical
history. A strangeness suppression factor $\lambda=0.5$ was adopted in
the Feynman-Field jet fragmentation model to address this observation
in experimental data~\cite{b13}. Other authors have presented evidence
for a value closer to $\lambda=0.3$~\cite{b14,b15}.

  There is probably nothing fundamental that one can learn from the
measured cross sections in Table \ref{sig_table} without correcting
them for secondary contributions that come from higher mass decays.
Nevertheless, we have examined the 6 elementary hadron cross sections
to see if our data might be made to appear more uniform with respect
to mass dependence by using a single strangeness suppression factor $\lambda$.
To some extent this effort will serve to compare our raw data 
to that of previous
experiments. We have explored 
3 variations of strangeness suppression.

 We first assumed that
if there were no strangeness suppression, all cross sections would be
described by 
\begin{equation}
\frac{d\sigma}{dy}\Big|_{y=0}=A e^{-Bm},
\label{simple_exp}
\end{equation}
\newline
where m is the particle rest mass. We then explored variations of 
strangeness suppression for the following 3 cases:
 (a) A single divisor
$\lambda$  raises all strange particle cross sections equally.
 (b) A single divisor 
$\lambda$  is applied to measured cross sections for $K^+$ and 
$\Lambda^0$ because
they contain only one strange quark, but a divisor of 
$\lambda ^2$ is applied to
$\phi$ and $\Xi^-$ cross sections because each of these particles contains
2 quarks with strangeness
quantum numbers. (c) A divisor of $\lambda $ is applied to $K^+$, $\Lambda^0$,
and $\phi$ cross sections because only one quark pair must be produced to 
form these particles, but a divisor of $\lambda ^2$ is applied to the
$\Xi^-$ cross section because two quark pairs must be produced to form it. 

  The data are fitted for values of A, B, and $\lambda$ that minimize
$\chi^2$. The values of $\chi^2$ per degree of freedom for each
of the cases listed above are: (a) $\chi^2_{dof}$=4.43, 
(b) $\chi^2_{dof}$=1.33, (c) $\chi^2_{dof}$=4.37. The best of these
fits is case (b), and the dashed line in Fig.~\ref{fs_only} shows  
graphically the extent to which the adjusted strange particle 
cross sections fit for $\lambda=0.437$. 

\begin{figure}[h]
\includegraphics[height=3.5in,angle=90]{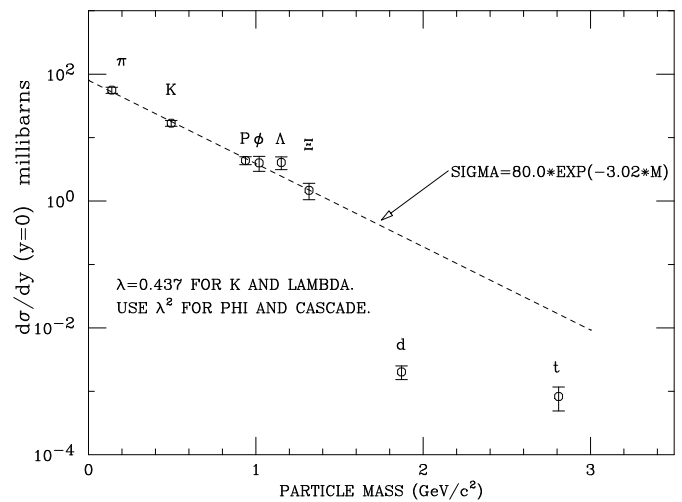}
\caption{\label{fs_only} Six inclusive hadron cross sections 
of Table~\ref{sig_table}
are plotted after multiplying $K^+$ and $\Lambda^0$ cross sections by
a factor $\lambda^{-1}$ and multiplying $\phi$ and $\Xi^-$ cross sections by a
factor $\lambda^{-2}$. The best fit to an exponential in mass
is the dashed curve for a strangenss suppression value
of $\lambda$=0.437 with a $\chi^2_{dof}$=1.33.}
\label{fs_only}
\end{figure}

  If we wish to attach some physical significance to the fit
such as interpreting the constant B as inverse temperature,
then we should use Eq.(\ref{spin_boltz}) below instead of 
Eq.(\ref{simple_exp})

\begin{equation}
\frac{d\sigma}{dy}\Big|_{y=0}=(2J+1)Ae^{-m/T},
\label{spin_boltz}
\end{equation}
\newline
where (2J+1) is the spin degeneracy of a particle
or resonance. To see exponential behavior one must
plot $[1/(2J+1)]d\sigma/dy|_{y=0}$. These values 
of cross section per spin state are given in Table~\ref{sig_spin}.

\begin{table}[h]
\caption{\label{sig_spin} Values of $d\sigma/dy$ at $y=0$ for hadronic
particles produced in this experiment by $p \bar p$ collisions
at $\sqrt{s}=1.8$ TeV. Spin degeneracy is removed in the last column.}
\begin{ruledtabular}
\begin{tabular}{cclc}
 Particle   &Mass (GeV)         & $d\sigma/dy|_{y=0}$        
& $d\sigma/dy|_{y=0}/(2J+1)$\\
\hline
 $\pi^+$    &0.139         &  56.0 $\pm$ 6.3 \rm{mb}        
& 56.0 $\pm$ 6.3 \rm{mb} \\
 $K^+$      &0.494         &   7.37 $\pm$ 0.79 \rm{mb}      
&  7.37 $\pm$ 0.79 \rm{mb}\\
 $p$        &0.938         &   4.37 $\pm$ 0.62 \rm{mb}      
& 2.19 $\pm$ 0.31 \rm{mb} \\
 $\phi$     &1.019         &   0.763 $\pm$ 0.202 \rm{mb}    
&  0.254 $\pm$ 0.067 \rm{mb}\\
 $\Lambda^0$&1.116         &   1.77 $\pm$ 0.41 \rm{mb}      
&  0.885 $\pm$ 0.203 \rm{mb}\\
 $\Xi^-$    &1.321         &   0.282 $\pm$ 0.082 \rm{mb}    
&  0.141 $\pm$ 0.041 \rm{mb}\\
 $d^+$      &1.871         &   2.02 $\pm$ 0.49 $\mu$b       
&  0.673 $\pm$ 0.163 $\mu$b\\ 
 $t^+$      &2.810         &   0.83 $\pm$ 0.34 $\mu$b       
&  0.415 $\pm$ 0.170 $\mu$b\\
\end{tabular}
\end{ruledtabular}
\end{table}

  The three cases for a strangeness suppression $\lambda$ that were
discussed above were also fit to the spin-scaled  cross sections
of Table~\ref{sig_spin}. The values of $\chi^2$ per degree of
freedom for these three cases were: (a) $\chi^2_{dof}=10.7$, 
(b) $\chi^2_{dof}=3.08$, (c) $\chi^2_{dof}=12.06$. The best
fit is once again for case (b), but the value of $\chi^2_{dof}$
is even less satisfactory than the fit to raw data.
For this case $\lambda =0.467$, and interpretation of the mass 
dependence implies $T=248 \pm 11$ MeV.

\section{Thermal Description including Particle Decays}

  The assumption of a strangeness suppression factor $\lambda$ 
improves the exponential appearance
of the data, but it does not return a statistically satisfactory
description of the measured data. Also it is not clear if $\lambda$ results
from particle production or particle decay or some combination
of the two processes.

  It is most certain that particle and resonance \emph{decays} favor
pion and proton final states over strange particle final states.
Every particle and resonance ultimately results in pions
and nucleons, whereas only flavored particles and flavored resonances
can decay into strange particles. 

  We have investigated a thermal model which has no strangeness
bias in the production of particles and resonances. It does, however,
use the actual strangeness bias found in decays as 
tabulated by the PDG~\cite{b16}
for 99 of the lowest mass particles and resonances.
The model assumes that a particle of given mass is produced
in $p\bar p$ collisions 
with equal probability in $(2J+1)$ spin states, $(2I+1)$
isospin states, and particle/antiparticle states. Each spin state 
for a specified charge is
assigned a primary production cross section per spin state according to its 
rest mass m as in
Eq.(\ref{prime_boltz}).

\begin{equation}
\frac{d \sigma^p}{dy}\Big|_{y=0}=A e^{-m/T}
\label{prime_boltz}
\end{equation}
\newline
  Observed particles are considered to result from either primary or
secondary production. Primary production cross sections
, $d \sigma^p/dy|_{y=0}$, must be estimated using
the primary particle mass in Eq.(\ref{prime_boltz}) since experimental
values for these cross sections are unknown. Secondary
contributions are determined by the primary production rates for some higher 
mass resonances
as given in Eq.(\ref{prime_boltz}) followed by a variety
of branching ratios in their ultimate decay to a primary particle
of interest. For low mass particles these branching ratios are
very well known and have errors that range from a fraction of a
percent to several percent~\cite{b16}.

  We are interested in the primary production of a particle
because only that rate might be expected to follow the exponential of
Eq.(\ref{prime_boltz}). Our goal is to learn what fraction
$x$ of the measured inclusive cross section for a particle is the result
of primary production. Measured cross sections $d \sigma/dy$ can then
be related to primary cross sections per spin state $d \sigma^p/dy$ by
Eq.(\ref{fp_sigma}).

\begin{equation}
(2J+1)\cdot \frac{d \sigma^p}{dy}\Big|_{y=0} =
 x \cdot \frac{d \sigma}{dy}\Big|_{y=0}
\label{fp_sigma}
\end{equation}
\newline

  The value we find for the fraction $x$ of a particular particle 
is a function of
the assumed temperature $T$ in Eq.(\ref{prime_boltz}). 
For example, when we try to determine the primary $\pi^+$ inclusive spectrum,
we may find that a massive resonance with large $J$ and $I$ values 
produces many $\pi^+$ mesons by its decay chains, but because its
resonance mass is much greater than $T$, it makes a relatively small
secondary contribution to the observed $\pi^+$ inclusive spectrum. 

  Branching ratios, conservation laws, and Clebsch-Gordon
coefficients were used to determine how much each primary state contributed
to observed $\pi^+$, $K^+$, p, $\phi$, $\Lambda^0$, and $\Xi^-$ 
inclusive spectra in the E735 detector. 
For particles with lifetimes of $\approx 10^{-10}$
seconds it was necessary to use a Monte Carlo program to evaluate average
corrections for loss of particles due to detector aperture and
analysis cuts. Typically $\sim$65\% of these protons were
accepted and $\sim$15\% of these pions. Strong decays give prompt
secondary particles at the beamline and were given no  relative corrections
for the detector losses. 

\begin{figure}[h]
\includegraphics[height=3.5in,angle=90]{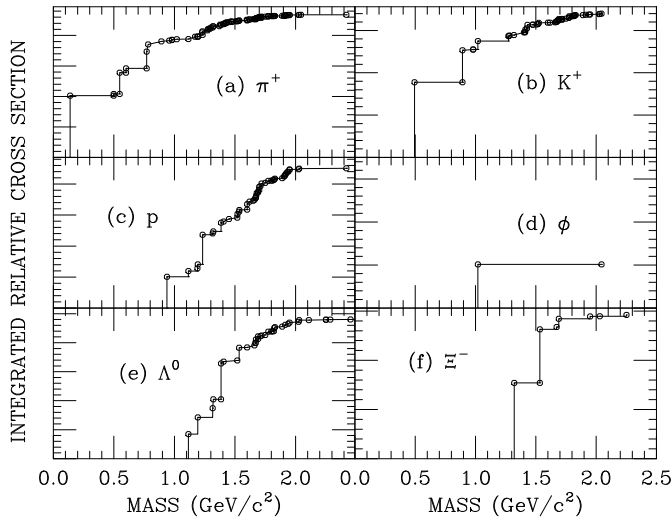}
\caption{\label{tableau_tst} 
 Six plots are shown for 
a common assumed value of $T$.
They display relative integrated cross section for a particle
as a function of the mass of contributing resonance or
particle decays. In general the limiting values of 
these curves will depend on the choice of $T$.
These plots use the value $T=204$ MeV, which minimizes the 
chi-square fit of primary cross sections per spin state
to Eq.(\ref{prime_boltz}).}
\end{figure}

   The tableau in Fig.~\ref{tableau_tst} shows, as a function 
of the mass of a
particle/resonance, the relative integrated contributions from various
resonances and particles to the 6 inclusive cross sections we are
studying. Vertical scales in the figure were chosen arbitrarily
in order to illustrate better the limiting contributions
from decays of higher masses. The exact shape and limiting values 
of these curves
will have a dependence on the choice of $T$ used in Eq.(\ref{prime_boltz}). 
The value of $T$ was chosen to be 204 MeV in order to produce 
the set of curves shown.
The ratio of the lowest mass step (primary production) to the limiting
integrated value yields $x$, the fraction that is primary production.

  A two parameter fit was done for $A$ and $T$ in Eq.(\ref{prime_boltz})
using the 6 inclusive cross sections per spin state in 
column 4 of Table~\ref{sig_spin}. In order to do this it was necessary
to make the functional dependence of the primary fraction, $x$, on T
available to the $\chi^2$ fitting program.
The fit is excellent with a $\chi^2$ per degree of freedom
$\chi^2_{dof}=0.317$ for 4 degrees of freedom. 

\begin{figure}[h]
\includegraphics[height=3.5in,angle=90]{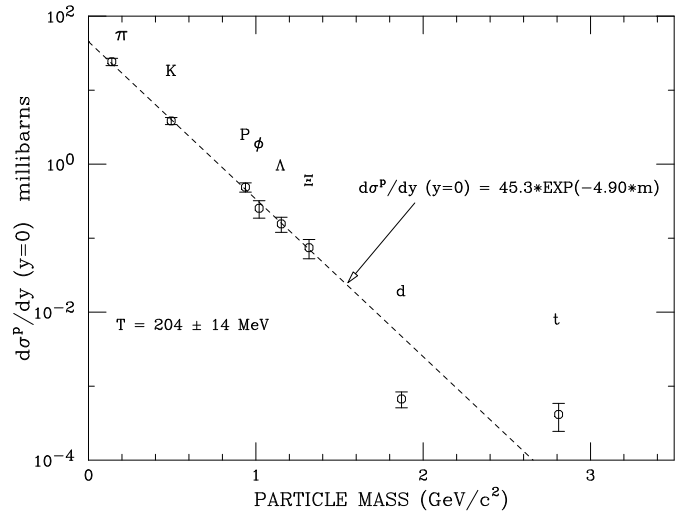}
\caption{\label{kt_expfit} Measured cross sections are scaled
down to remove secondary contributions
from decay of higher mass resonances and then divided by 
spin degeneracy $(2J+1)$ before plotting. The dotted line
is a Boltzmann distribution in mass using $T=204$ Mev obtained from
a fit to the primary cross sections for the 6 elementary 
particles. The two nuclear cross sections per spin state are
plotted to show the extent to which they do not
appear to participate directly in this simple thermal model.}
\end{figure}

  Table~\ref{fp_table} lists for each measured particle the value of x
and the primary cross section per spin state as determined by the
fit. The primary cross sections per spin state are
plotted in Figure~\ref{kt_expfit} along with the fitted function
(dashed curve). The temperature determined by the fit is
$T=204 \pm 14$ MeV. This is significantly higher than 
$T=160$ MeV suggested for the bootstrap statistical model~\cite{b17}
and quite close to $T=200$ MeV suggested for the QGP~\cite{b18}.

\begin{table}[h]
\caption{\label{fp_table} Values of $d\sigma/dy$ at $y=0$ for hadronic
particles produced in $p \bar p$ collisions
at $\sqrt{s}=1.8$ TeV. Column 2 lists cross sections with spin degeneracy
removed. Column 3 lists factors by which column 2 must be multiplied to
obtain primary cross sections per spin state listed in column 4.}
\begin{ruledtabular}
\begin{tabular}{cccc}
 Particle           & $d\sigma/dy|_{y=0}/(2J+1)$     &  $x$   
& $d\sigma^p/dy|_{y=0}$(mb) \\
\hline
 $\pi^+$            & 56.0 $\pm$ 6.3 \rm{mb}       &  0.433   
&   24.3 $\pm$ 2.7        \\
 $K^+$              &  7.37 $\pm$ 0.79 \rm{mb}     &  0.522   
&    3.85 $\pm$ 0.41      \\
 $p$                &  2.19 $\pm$ 0.31 \rm{mb}     &  0.223   
&    0.489 $\pm$ 0.069    \\
 $\phi$             &  0.254 $\pm$ 0.067 \rm{mb}   &  0.999   
&    0.254 $\pm$ 0.067    \\
 $\Lambda^0$        &  0.885 $\pm$ 0.203 \rm{mb}   &  0.176   
&    0.156 $\pm$ 0.036    \\
 $\Xi^-$            &  0.141 $\pm$ 0.041 \rm{mb}   &  0.527   
&    0.0743 $\pm$ 0.0216  \\
\end{tabular}
\end{ruledtabular}
\end{table}

  The extraordinarily good fit to a simple exponential
in mass is somewhat of a surprise since it relies in
no way on the significant amount of translational energy
observed experimentally for these particles.  
There is also no hint
of a distinction between Bose-Einstein statistics and 
Fermi-Dirac statistics. We do have good evidence that
there is some Bose-Einstein interference between observed
pion pairs in this experiment~\cite{b19}.

  The best fit values for A ($45.3 \pm 2.8$ mb)
and T ($204 \pm 14$ MeV) are remarkable for their apparent
relation to other meaningful physical parameters. As we note in 
Section XV , if cross sections fall exponentially, the 
Uncertainty Principle suggests that
$T=\hbar c/r_h$, where the hadronic radius is
$r_h \approx 0.97 \pm 0.07$ fm.  The value of 
$A=45.3 \, \mathrm{mb} \approx 0.721 \cdot (\pi \lambdabar_{\pi}^2)$
should be compared to the fact that the $\pi^{\pm} p$ cross sections~\cite{b16}
at the J=3/2, I=3/2 resonance saturate for a value given by
\begin{equation}
\label{fermi}
\sigma(\pi p) \approx 0.797 \cdot (\pi \lambdabar_{\pi}^2)(2J+1).
\end{equation}

\section{Extrapolation to Higher Masses}

  If particle production as described by the above thermal model 
is valid for higher particle masses, 
then the thermal model 
can serve as a lower limit for presently unmeasured cross sections or
it can be used as a standard measure for identifying unusually large
cross sections. 

  Checking the validity of the thermal model at high mass values is
difficult for two main reasons. One reason is that almost all event triggers
at the large collider detectors rely on finding particles with high
$p_t$. The few cross sections which have been measured apply only to
the highest portion of some incompletely known momentum spectrum.

  Another reason is that detailed branching ratios like those we
used to prepare Fig.~\ref{tableau_tst} are unknown or unpublished, so
that one cannot easily calculate a primary cross section. 
Only if one could find a
particle like the $\phi$ meson (Fig. 3d) with 
negligible contributions from nearby mass resonances, would it be
unnecessary to know detailed branching ratios for that particle.

\begin{figure}[h]
\includegraphics[height=3.5in,angle=90]{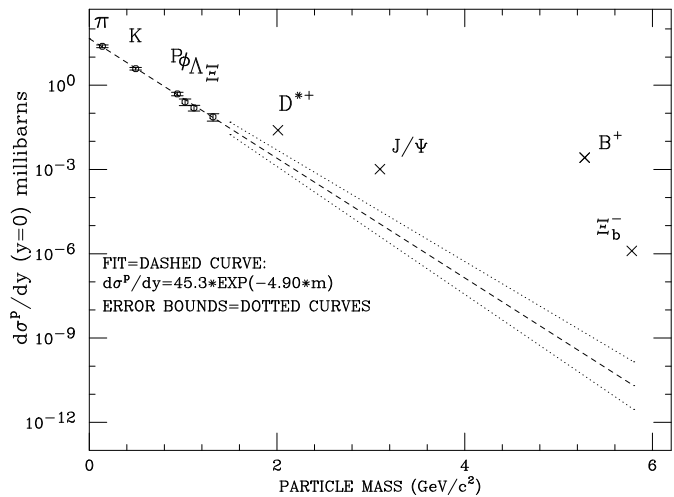}
\caption{\label{predict} The dashed curve is a fit to
cross sections observed in this experiment. In the figure
the curve is extrapolated to much higher mass values and
is bounded by dotted curves which represent statistical
errors for the fitted temperature.}
\end{figure}

  We have attempted to make order-of-magnitude cross section estimates
per spin state
for four massive particles using reported data. These are plotted as
X's without error bars in Fig.~\ref{predict} for comparison with the
extrapolated thermal model. Some calculational details of these estimates
are outlined in the Appendix. 

  An extrapolation of the fit to our six primary cross sections is shown as
a dashed line in Fig.~\ref{predict} bounded by dotted lines representing the
slope errors for the fit. Although the higher mass cross section 
estimates in this
figure have varying degrees of unreliability, they all lie 
above the expectations
of thermal production, and the discrepancy tends to grow with increasing mass.

  The $D^{*+}$ cross section appears to be almost an order of magnitude above
the thermal production expectations. There are very few examples of secondary
decay contributions from nearby resonance decays into $D^{*+}$ that are listed
by the PDG~\cite{b16}. Admixtures of B mesons are expected to decay as
$\, B^{\pm o} \rightarrow D^{*+} + Anything \,$ 22.5\% of the time~\cite{b16},
but the $B^+$ cross section alone is about an order of magnitude below the 
$D^{*+}$ cross section. 

   If all secondary sources of $D^{*+}$ can be accurately accounted for, 
it may
eventually be possible to calculate a primary cross section 
that agrees with the
thermal model. However, since the number of secondary sources tends to
increase exponentially with mass~\cite{HagIII}, it may never be practical
to isolate primary cross sections for the heavier particles.  

   The estimated $J/\psi$ cross section is approximately 
two orders of magnitude above
the thermal model prediction. There are 11 spin states of $\chi_{c1}$, 
$\chi_{c2}$, and
$\psi(2S)$ which have significant decays (20\% to 56\%) into $J/\psi$. 
However, it
seems unlikely that the primary thermal production of $J/\psi$ is
a source which is competitive with other
constituent scattering sources.

   Momentum distributions have been measured for $D^{*+}$, $J/\psi$, and
$B^+$ production perpendicular to
the colliding beams~\cite{b23,b22,b27}. In the range of $6<p_t<20$ GeV/c 
the $D^{*+}$ and $J/\psi$ spectra
fall off with a definite inverse power law ($d \sigma/dp_t\sim 1/p_t^5$). 
This
is the same large $p_t$ dependence observed in pion production at $90^0$ 
for large 
$\sqrt{s}$ collisions~\cite{b34}. Independence of $\sqrt{s}$ and particle 
mass are
sufficient reasons to identify this common behavior as 
Feynman-scaling~\cite{b35}, which is
typically associated with hard parton scattering~\cite{b36}. 
However, the $p_t$ dependence of the
cross sections we have measured in this experiment 
is better represented by an exponential function 
up to 1.5 GeV/c for identified
protons and up to 3.0 GeV/c for $\Lambda^o$'s.
At large particle masses where
the thermal cross sections are expected to be very small, hard parton scattering
may be the dominant production mechanism.

\section{Hadrons from $\,\mathbf{e^+e^-}$ Collisions}

  Although reliable cross section values for higher mass particles
are not available for $p\bar p$ collisions, some have been measured in $e^+e^-$
collisions. It would be interesting to learn if the thermal model in some
way mediates hadron production in electron collisions. We have chosen
to investigate this possibility by studying multiplicity fractions
for charm particle production. 

  The multiplicity fraction $f_h$ for a particle is the
ratio of the inclusive cross section for producing the particle to the total
cross section for producing hadrons: $f_h=\sigma^{incl}/\sigma^{hadron}$.
Thus we treat multiplicity fractions for charm production
in $e^+e^-$ collisions in the same way as we treat inclusive charm
cross sections resulting from hadron collisions.
For example, the multiplicity fraction for $D^{+}$ production, $f_h(D^+)$,
contains a primary component as well as secondary contributions from the
decay of higher mass particles.

\begin{figure}[h]
\includegraphics[height=3.5in,angle=90]{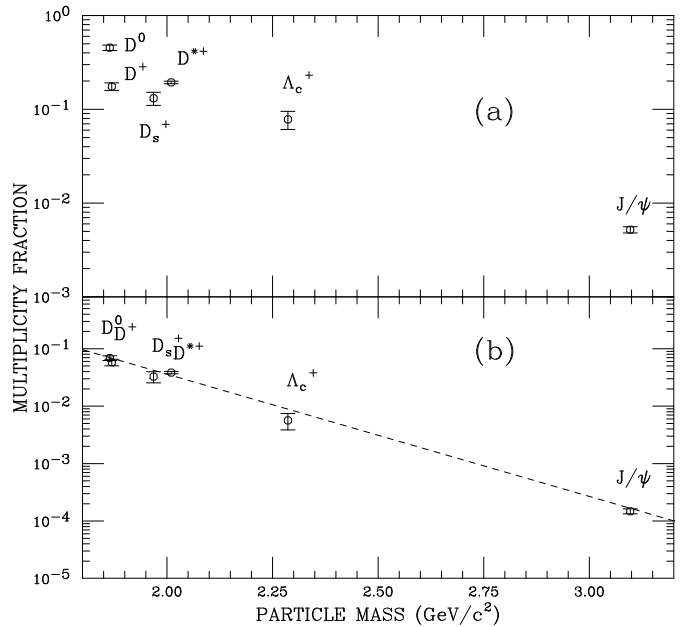}
\caption{\label{fract} (a) Observed hadronic multiplicity fractions at
$\sqrt{s}=91$ GeV~\cite{b25}
vs. charm particle masses. (b) Primary multiplicity fractions per spin state 
after correcting
for secondary decay contributions. Dashed curve is an exponential
$6.53 \times 10^2\,e^{-m/T}$, where m is particle mass and $T=0.204$ GeV.}
\end{figure}

  The PDG provides tables~\cite{b25} of the total hadronic fractions $f_h$
for a set of charm particles. Although data points are 
tabulated for collisions
at $\sqrt{s}=91$ GeV and $\sqrt{s}=10$ GeV, we will focus on the data from
$\sqrt{s}=91$ GeV in an effort to achieve smaller errors. 
Values of $f_h$ measured at $\sqrt{s}=91$ GeV are listed for 6 charm particles
in Table~\ref{frag_frac_c} and plotted in Fig.~\ref{fract}(a).

  Because the production rates of b and c quarks in $e^+e^-$ scattering 
are comparable
in size at $\sqrt{s}=91$ GeV (the $Z^0$ mass), the contribution of b quarks 
to the total hadronic 
fraction of a charm particle can be substantial and must be subtracted in order 
to obtain the fraction
contributed by c quarks, 

\begin{equation}
  f_c=f_h-f_b . \label{fh-fb}
\end{equation}

   Fortunately it is possible to obtain $f_b$ 
for a limited number of charm particles from a set of partial widths $\Gamma$
tabulated by the PDG.~\cite{b39}
Using the partial width 
$\Gamma (b \rightarrow c)$ for a b-admixture to yield a charm particle c,
one can calculate $f_b$ directly.

\begin{equation}
 f_b=\Gamma (b \rightarrow c) \times \Gamma(Z^0 \rightarrow b \bar b)/
\Gamma (Z^0 \rightarrow hadrons)\label{fb1}
\end{equation}

\begin{equation}
 f_b=\Gamma (b \rightarrow c) \times (0.2163 \pm 0.0007), \label{fb2}
\end{equation}
\newline
where $\Gamma(Z^0 \rightarrow b \bar b)$ and $\Gamma(Z^0 \rightarrow hadrons)$
are the partial widths for $Z^0$ decay into 
$b \bar b$ and hadrons respectively~\cite{b40}.

  The first five values of $f_c$ listed in Table~\ref{frag_frac_c}
were obtained as outlined above. Errors for these values range from $5\%$ to
$30\%$. However, the last value, which is for $J/\psi$ production, 
presents a special case
because $~95\%$ of $J/\psi$ particles produced at $\sqrt{s}=91$ GeV 
come from the decay of B hadrons~\cite{OPAL}. The error on the difference in 
Eq.(\ref{fh-fb}) may be comparable to the difference itself.

  One might consider associating $f_c$ with the experimentally 
prompt component of $J/\psi$ production
at $\sqrt{s}=91$ GeV, but the statistical accuracy on a direct measurement
of prompt $J/\psi$ production at this energy is poor~\cite{OPAL}.  
Nevertheless, the
prompt value at $\sqrt{s}=91$ GeV is close to the 
total value~\cite{b25} obtained
with much greater accuracy at $\sqrt{s}=10$ GeV 
(where b quarks are not produced).
We have used this more accurate value for $f_c$ in Table~\ref{frag_frac_c} for
$J/\psi$ production.

\begin{table}[h]
\caption{\label{frag_frac_c}This table lists multiplicity fractions extracted
from charm production in $e^+e^-$ collisions~\cite{b25}. Column 1 identifies
the charm particle. Column 2 gives the experimentally observed fractions
for charm particle production at $\sqrt{s}=91$ GeV. Column 3 gives the fraction
when contributions from b decays are removed. 
Column 4 gives the primary fraction
after contributions from higher mass charm decays are also removed.}
\begin{ruledtabular}
\begin{tabular}{cccc}
Name & $f_h$ & $f_c$ & $f_c^p$ \\
\hline
$ D^0 $ & 0.454 $\pm$ 0.030 & 0.325 $\pm$ 0.031 & 0.0690  $\pm$ 0.0065\\
$ D^+ $ & 0.175 $\pm$ 0.016 & 0.125 $\pm$ 0.016 & 0.058   $\pm$ 0.008\\
$ D_s^+$ & 0.131 $\pm$ 0.021 & 0.099 $\pm$ 0.022 & 0.033   $\pm$ 0.007\\
$ D^{*+}$ & 0.194 $\pm$ 0.006 & 0.156 $\pm$ 0.007 & 0.0384  $\pm$ 0.0017\\
$ \Lambda_c^+$ & 0.078 $\pm$ 0.017 & 0.057 $\pm$ 0.018 & 0.0057  
$\pm$ 0.0018 \\
$ J/\psi$ &0.0052 $\pm$ 0.0004 & 0.0005 $\pm$ 0.00005 & 0.000148 
$\pm$ 0.000015 \\

\end{tabular}
\end{ruledtabular}
\end{table}

  Our objective is to determine if the 6 observed values of $f_h$ 
in Table~\ref{frag_frac_c}
are consistent with the thermal model that describes hadron production 
in $p \bar p$ 
collisions. To do this we must correct $f_c$ for contributions from
secondaries of other charm decays. We can write $f_c$ as a sum 
of contributions as follows.

\begin{equation}
\mathop{ f_c=(2J_c+1)f_c^p + \sum_{i}
(2J_i+1) f_i^p \mathfrak{B_{ic} } }
\label{fobs}
\end{equation}

  In this expression $f_c$ is the observed multiplicity fraction 
for charmed particle c. The
primary fraction per spin state for particle c is $f^p_c$, and $J_c$ is the spin
of particle c. Similarly $f^p_i$ is the primary fraction of particle i with spin
$J_i$ that contributes secondary decay products  to the  multiplicity fraction
observed for particle c. $\mathfrak{B_{ic}}$
is the branching fraction of particle $i$ into particle c, and in some cases
it is the result of a decay chain.

 In the charm sector the branching fractions $\mathfrak{B_{ic}}$ are sometimes
poorly determined. We make use of data summarized in 2008 by the PDG~\cite{b16}.
For some particles branching fractions into hadrons
are given as ``seen'' or ``dominant''. In order to obtain numerical estimates
in these cases we make use of isospin symmetry, phase space ratios, and 
Clebsch-Gordon coefficients. Usually the most uncertain branching fractions
are for higher mass particles which have smaller multiplicity
fractions and thus
contribute fewer secondaries.  

To test for consistency we estimate the primary fragmentation $f_i^p$ using
the exponential fall off which we found for cross sections in 
$p \bar p$ collisions,
\begin{equation}
\mathop{f_i^p=f_c^p e^{-(m_i-m_c)/T}=f_c^p \epsilon_{ic}},
\label{f_alpha_i}
\end{equation}
\newline
where $T=0.204$ GeV. After inserting this in Eq.(\ref{fobs}),
we can solve for $f_c^p$, the primary fragmentation fraction 
per spin state of charmed particle c.
\begin{equation}
\mathop{f_c^p=
\frac{\displaystyle f_c}
{\displaystyle \biggl[
(2J_c + 1)+\sum_{i} (2J_i + 1)\mathfrak{B_{ic}}\epsilon_{ic}\biggr] } }
\label{f_alpha}
\end{equation}
\newline
This expression has two useful attributes. (1) Computation of $f_c^p$ 
depends only on 
the measurement of one multiplicity fraction,
$f_c$, and not on other measured fragmentation fractions and 
their possible experimental
errors. This could be useful in identifying a single faulty 
measurement since there is no 
interdependency on other multiplicity measurements. 
(2) The primary fraction $f_c^p$ can
never be negative, which can happen if one ``subtracts'' secondary contributions
using other measured fractions that have experimental errors and uncertain
double counting in decay chains. 

\begin{table}[h]
\caption{\label{corr_sources} Column 2 of this table lists sources of
secondary particles that were used to correct the observed fractions
for the particles in column 1.}
\begin{ruledtabular}
\begin{tabular}{cc}
Particle& Secondary Sources \\
\hline
$D$         & $D^*,D_0(2400),D_1(2420),D_1(2430)$\\
            & $D_2(2460),D_1(2640),D_{s1}(2536),D_{s2}(2573)$\\
$D_s$       & $D_s^*(2112),D_{s0}(2317),D_{s1}(2460),D_{s1}(2536) $\\
$D^*$       & $D_1(2420),D_1(2430),D_2(2460)$\\
            & $D_1(2640),D_{S1}(2536)$\\
$\Lambda_c$ & $\Sigma_c (2455),\Sigma_c (2520),\Lambda_c (2595),
\Lambda_c (2625) $\\
            & $\Sigma_c (2800),\Lambda_c (2880) $\\
$J/\psi$      & $\chi_{c1}, \chi_{c2}, \psi(2S), X(3872) $ \\
\end{tabular}
\end{ruledtabular}
\end{table}

   The sources used for secondary decay contributions are listed explicitly
in Table~\ref{corr_sources}. All of these sources have been limited to the
charm sector since the earlier subtraction of $f_b$ accounted for the 
contributions from B hadrons.

  Six calculated primary multiplicity fractions per spin state, $f_c^p$, 
are listed in column 4 of
Table~\ref{frag_frac_c} and plotted in Fig.~\ref{fract}(b). 
These were computed using
Eq.(\ref{f_alpha}) with $f_c$ data from column 3 of Table~\ref{frag_frac_c}. 
The dashed line
is a fit for normalization using $f_c^p=A\,e^{-m/0.204}$, which gives 
$A=653 \pm 24$ with $\chi^2_{dof}=2.8$. This could be considered a
reasonably good fit, since the error bars in  
Fig.~\ref{fract}(b) simply employ
the same percentages as those
given for the observed multiplicity fractions, and no allowance is made
for systematic inconsistencies among various experiments.  

  When corrections for all secondary decay contributions are completed,
the $D^0/D^+$ isospin violation evident in Fig.~\ref{fract}(a) disappears 
in Fig.~\ref{fract}(b), which is encouraging
since we did not use directly measured $D^*$ fractions to bring about the
agreement. This isospin symmetry violation 
(difference in multiplicity fractions)
for the $D^+$ and $D^0$ is understood in terms
of energy violation for decay of $D^*$ into $D^+$~\cite{b28}. 

   One can conclude that the six charm multiplicity fractions 
in Fig.~\ref{fract}(b)
are consistent with a thermal production model. 
We have optimistically assumed that
errors in the branching 
fractions $\mathfrak{B_{ic}}$ are negligible. Since branching ratios for 
decays of higher mass particles are sometimes poorly determined, these errors
in some cases could be significant.  

   We have also attempted this same analysis using $e^+e^-$ multiplicity 
fractions tabulated
entirely for $\sqrt{s} \approx 10.5$ GeV~\cite{b25}. 
By working at that energy experimenters 
should have eliminated
b-quark fragmentation from the data. Our analysis result
at this lower value of $\sqrt{s}$ yields a fairly good
description by $T=204$ MeV for  the non-strange charm particles, 
but there is
a ``strangeness suppression'' of about a factor of two 
for $D_s^+$ and $D_s^{*+}$
mesons which does not appear at $\sqrt{s}=91$ GeV. This could be a physical
effect that is related to the choice of a lower $\sqrt{s}$ or it
could be the result of some systematic experimental error, such as a slightly
incorrect Monte Carlo model.   

\section{Models for Primary Cross Sections}

The scale factor for our cross sections $(\pi \lambdabar_{\pi}^2)$ has been
a familiar feature since the discovery of the $\Delta (3,3)$ resonance, and the
temperature (T) appears to be almost a trivial consequence of hadronic size
and the Uncertainty Principle. Nevertheless, one would like to find a more
fundamental description of particle production in high energy collisions that
could give physical meaning to these and other observables.

  In what follows we will examine several descriptions. The foremost
requirement for each of these is that it must
be compatible with the observed primary cross sections.
In addition to that, a useful description should 
guide us to some reasonable approximation
for the measured average transverse momentum $<\!\!p_t\!\!>$ of each
studied particle. The {\it measured} value of $<\!\!p_t\!\!>$ is,
of course, not that of the particles from primary production, since
there is in general no experimental way to distinguish primaries from
the secondaries of resonance decays. 
One can hope that the random nature of the decay process will have less 
effect on the average $p_t$  than it will on the exact shape of
the $p_t$ distribution. We present some Monte Carlo calculations
in Section XI that support this hope. Finally, a successful description 
should yield some
gross approximation to the observed $p_t$ spectra after making
reasonable allowances for distortions by contributions from secondaries
of resonance decays.

 As we investigate descriptions, we will recognize the possibility that the
final state of the particles may be different from the initial 
production state. That is, we consider cases where the final 
temperature may be different from the initial temperature as
long as the particle ratios remain the same. 

\subsection{Hawking-Unruh Radiation}

 A recent approach to particle production has been to consider it as a
consequence of Hawking-Unruh radiation~\cite{Kharzeev}.
In this description the rate R for the production of a particle of mass m is
\begin{equation}
R \propto e^{-m/a} \qquad (\hbar=c=1),
\label{Hawk_eq}
\end{equation}
where a is the particle
acceleration. This is the result of a quantum mechanical 
tunneling calculation first
used by Schwinger to describe pair production by an electric 
field~\cite{Schwinger}. Hawking later used this method to calculate
the temperature of black holes, where the acceleration at the
event horizon was considered to serve exactly as a 
temperature~\cite{Hawking}.

When applied to high energy experiments, the acceleration process
is initiated by the particle collisions. Some estimates for the 
temperature are $T=Q_s/2\pi \approx 200$ MeV~\cite{Kharzeev}, 
$T=\sqrt{3 \sigma/4 \pi}$~\cite{T2}, $T=\sqrt{\sigma/2 \pi}$~\cite{T2}.
$Q_s$ is the ``saturation scale'' in the gluon saturation regime.
In the last two expressions $\sigma$ is the string tension 
between quarks in units of $MeV^2$. Our fit to the data with
$T=204$ MeV is in agreement with acceptable values of string tension. 

 It is important to note that the above estimates of acceleration do
not have a mass dependence. A more recent version of this 
approach~\cite{Tmass} introduces quark masses. The resulting
temperatures then imply strangeness and charm suppression.
In earlier sections we have shown there is no significant evidence 
to support suppression of strangeness or charm production in primary
cross sections.

\subsection{A Bootstrap Description}

  The bootstrap model was proposed by Hagedorn~\cite{HagIII}
after he first noted that the density of mass states known at that
time increased almost exponentially with mass. A single state is
defined as a particle or antiparticle with  particular 
eigenvalues for mass,
spin ($J_z$), and isospin ($I_z$). Frautschi~\cite{Frautschi} has written a
paper that refines the bootstrap assumptions and concisely
explains such a model and its consequences. The principal results are 
a limiting temperature $T_0$, sometimes called ``the boiling point
of matter'', and a limiting analytic
form for the density of states, $\rho (m)$, as a function of the mass m.
\begin{equation}
\rho (m) = c\,m^a\,e^{m/T_0} \qquad (a< -5/2).
\label{rho1}
\end{equation}
Hagedorn has suggested a ``smoothed'' version of this expression
which could be used at low masses or even at $m=0$~\cite{HagIII,Hag4},
\begin{equation}
 \rho (m) = c\, T_0^{3/2}\,(m_0^2+m^2)^{-5/4}\,e^{m/T_0}.
\label{rho2}
\end{equation}
Hagedorn expected values of $c=5.5^{3/2}$, $m_0=500$ MeV, and
$T_0=160$ MeV.

The inverse of this expression is so similar to the mass dependence of
$d\sigma/dy|_{y=0}$ that we have tried fitting the cross section data with
\begin{equation}
\frac{d\sigma^p}{dy}\Big|_{y=0}=A \, T^{-3/2} \, (m_0^2 +m^2)^{5/4} \, e^{-m/T},
\label{fit_rho}
\end{equation}
where A and T are fittable parameters, and $m_0=500$ MeV is Hagedorn's 
choice of a smoothing parameter.

\begin{figure}[h]
\includegraphics[height=3.5in,angle=90]{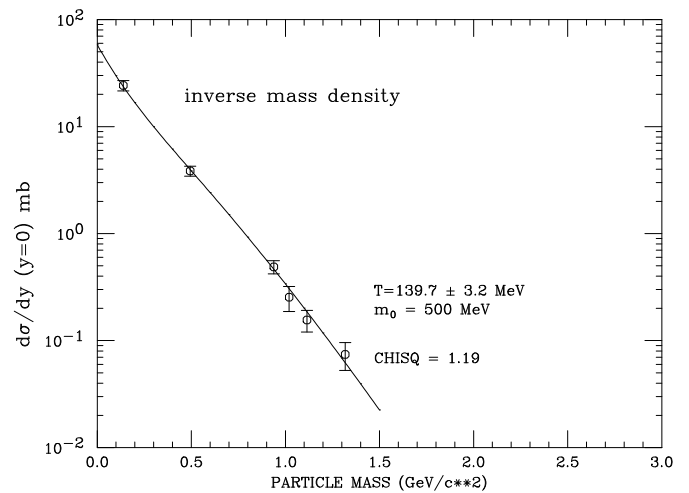}
\caption{\label{bootfit} The solid curve is a fit to
cross sections observed in this experiment  
using the inverse of the mass density $\rho (m)$ suggested
by the Hagedorn bootstrap model.}
\end{figure}

The results are shown in Fig.~\ref{bootfit} with
$A=17.2\pm 1.5$ mb/GeV, $T=0.1397 \pm 0.0032$ GeV, and
$\chi^2_{dof}=0.30$. The parameters A and $m_0$ are strongly
correlated and poorly constrained by the cross section
data, which is why we accept Hagedorn's determination for
$m_0$ without making it a free fitting parameter. 
We observe that $T_0>T$, so that the cross section
appears to decrease with increasing mass faster than the density of states
increases. 

 The use of this expression for the cross section suggests that to some extent
we may observe an exponential decrease in $d \sigma^p/dy$ with increasing
mass because we choose to measure only one of the many available 
states at mass m, whereas the number of available states increases 
exponentially with increasing mass.

\subsection{A Gas Model}

  In this section we will explore the possibility that a relativistic
gas model may describe the observed particle production cross sections.
We make use of an expression for the number of particles per unit volume,
$dn$, in the center of mass system given by~\cite{degroot}

\begin{equation}
 dn \propto \frac{d^3{\bf p}}{[e^{(E-\mu)/T} \mp 1]},
\label{BE_eqn}
\end{equation}
where $T$ is the gas temperature and $\mu$ is a chemical potential.
The volume element in momentum space, $d^3{\bf p}$, is given by

\begin{eqnarray}
d^3{\bf p} &=& 2\pi \, p_tdp_t\,dp_l \nonumber \\
           &=& 2\pi \, p_tdp_t \, E \, (\frac{dp_l}{E}) \nonumber \\
           &=& 2\pi \, p_t \, E \, dp_t dy
\label{ph_spc}
\end{eqnarray}

The total energy of a particle is E.  $p_t$ and $p_l$ are 
respectively components
of its momentum transverse and parallel to the beam
direction in the center of momentum system. The variable 
$y=(1/2)ln[(E+p_l)/(E-p_l)]$ is the rapidity. In our experiment
$p_l \approx 0$. Thus one can use
$E=\sqrt{p_t^2 + m^2}$ for the total energy of a particle of mass m
and take $y=0$. 

 In order to calculate a primary cross section a simple assumption is
made that
\begin{equation}
\frac{d\sigma^p}{dy}\Big{|}_{y=0}=A\frac{dn}{dy}\Big{|}_{y=0}.
\label{sig_eqn}
\end{equation}
To obtain the cross section for any particle it then becomes necessary
to evaluate the integral

\begin{equation}
\frac{dn}{dy}\Big{|}_{y=0} = \int_0^{\infty} 
\frac{p_t \, E \, dp_t}{[e^{(E-\mu)/T}\mp 1]}.
\label{int_eqn}
\end{equation}
 
These integrals are done numerically.
The second term in the denominator of the integrand should be -1
for bosons and +1 for fermions. 
In our applications we find that for masses above those of the kaon
and pion the second term is insignificant, so that all of the
calculations in this paper have actually used the boson form.

\begin{figure}[h]
\includegraphics[height=3.5in,angle=90]{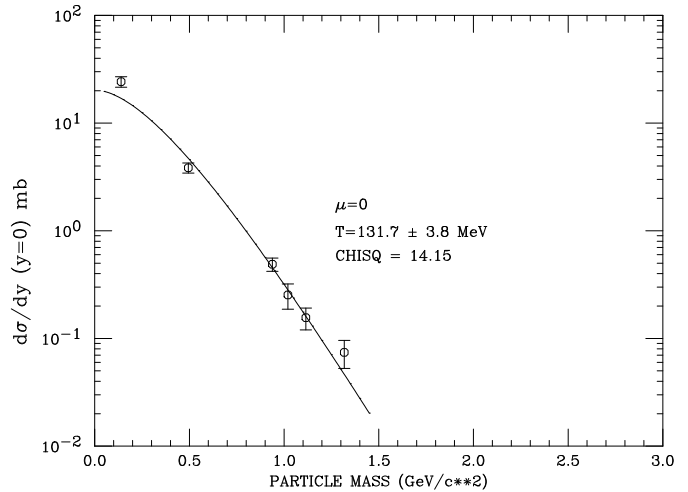}
\caption{\label{zero_mu} The curve is a fit of the statistical
gas model in Eq.(\ref{int_eqn}) to the primary cross sections
per spin state of 6 particles
when the chemical potential $\mu=0$ for each particle.}
\end{figure}

  A first attempt to describe the primary cross sections of Table~\ref{fp_table}
was made by assuming the chemical  potential $\mu=0$ for all particles
and by fitting equations \ref{sig_eqn} and \ref{int_eqn} to constants
A and $T$. The result is the curve shown in Fig.~\ref{zero_mu}.

Fitted values are $A=(3.69\pm 0.71) \times 10^3\quad \mathrm{mb/GeV}^3$ and
$T=0.1317 \pm 0.0038\quad \mathrm{GeV}$. Unfortunately the $\chi^2$
for this fit, $\chi^2=14.15$ for 4 degrees of freedom,
is unacceptable. The main difficulty seems to be a failure
to predict a large enough pion cross section. This difficulty
can be addressed by selecting a suitable chemical potential.

From the thermodynamics of a system with k types of constituents,
one knows that a change dU in the internal energy of the system
can be written as~\cite{reif}
\begin{equation}
 dU=TdS-pdV+\sum_{i=1}^k \mu_idN_i.
\label{thermo}
\end{equation}

It should be possible to estimate a value for the chemical potential
of a pion, $\mu_{\pi}$, using a gedanken experiment which adds only one 
pion ($dN_{\pi}=1$) to a system with temperature T and pressure p while
arranging for entropy S and volume V to remain constant ($dS=dV=0$). Then
the chemical potential has a value $\mu_{\pi}=dU$. 
This suggests that $\mu_{\pi}=m_{\pi}$, but the 
relation with the pion mass must be such that $\mu_{\pi} < m_{\pi}$ if 
the integrand of Eq.(\ref{int_eqn}) is to have no singular values. 

Provided the system already contains constituent u and $\bar d$ quarks,
one can imagine that dU need not supply the energy for existing 
quark masses. In that case we subtract existing quark masses~\cite{b16}
as below.
\begin{eqnarray}
 \mu_{\pi}=m_{\pi}-m_u -m_{\bar d} &=& 0.1396-0.010 \nonumber \\
                         \mu_{\pi} &=& 0.1296 \quad \mathrm{GeV}
\label{mu_value}
\end{eqnarray}

\begin{figure}[h]
\includegraphics[height=3.5in,angle=90]{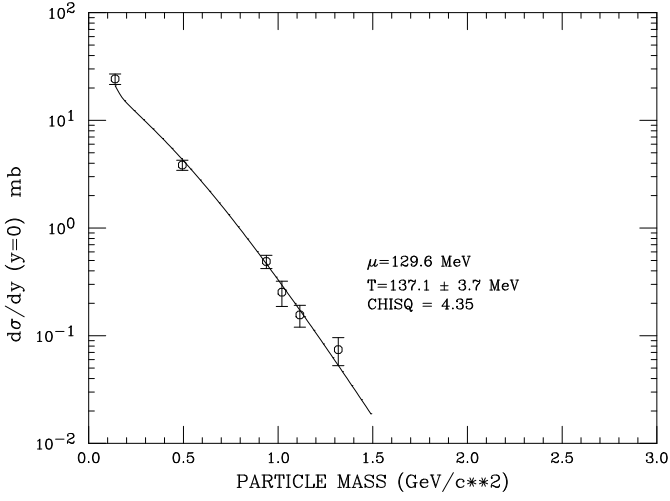}
\caption{\label{quark_mu} The curve is a fit of the
statistical gas model in Eq.(\ref{int_eqn}) to primary cross sections per
spin state when the chemical potential
$\mu=129.6$ MeV for each of the 6 particles.}
\end{figure}

We have used this same value of $\mu$ in Eq.(\ref{int_eqn}) for
each of the 6 measured primary cross sections in order to fit for new
values of A and T. This fitted curve can be compared to the data
in Fig.~\ref{quark_mu}.

The value of $A=(1.054 \pm 0.156) \times 10^3$ mb/Ge$\mathrm{V}^3$ and
$T=0.1371 \pm 0.0037$ GeV. The $\chi^2=4.35$ for 4 degrees of freedom
is quite satisfactory from a statistical point of view. 

  The use of a common chemical potential $\mu$ for all the particles 
works best for fitting the pion data, and as we shall see later,
it may influence the average kaon momentum in a beneficial way. It is
not possible with the present data set to exclude the possibility
that $\mu=0$ for the heavier particles. On the other hand, it is
easy to rule out the use of the thermodynamic arguments like
Eq.(\ref{thermo}) and Eq.(\ref{mu_value})  to determine a unique chemical 
potential for each mass because no reasonable fit to the
cross sections was possible for that approach. 

\section{Average Transverse Momenta}

Two significantly different temperatures are candidates for
describing the production cross sections we have measured,
T=204 MeV used with a simple exponential in the mass or
T=137.1 MeV used in a relativistic gas model. We will first try
to determine if either of these two temperatures is more
appropriate for characterizing the measured values of 
average transverse momentum. 

  Consulting Eq.(\ref{int_eqn}), we find that 
the relative transverse momentum spectrum
at y=0 should be given by

\begin{equation}
 \frac{dn}{dy\,dp_t}\Big{|}_{y=0}=\frac{p_t \, E}{[e^{(E-\mu)/T} \mp 1]}.
\label{diff_eqn}
\end{equation}

The average transverse momentum is then calculated for each mass value as

\begin{equation}
<\!\!p_t\!\!>=\frac{\int_0^\infty p_t (dn/dy \, dp_t) \, dp_t}
{\int_0^\infty (dn/dy \, dp_t) \, dp_t}.
\label{avg_pt}
\end{equation}

\begin{figure}[h]
\includegraphics[height=3.5in,angle=90]{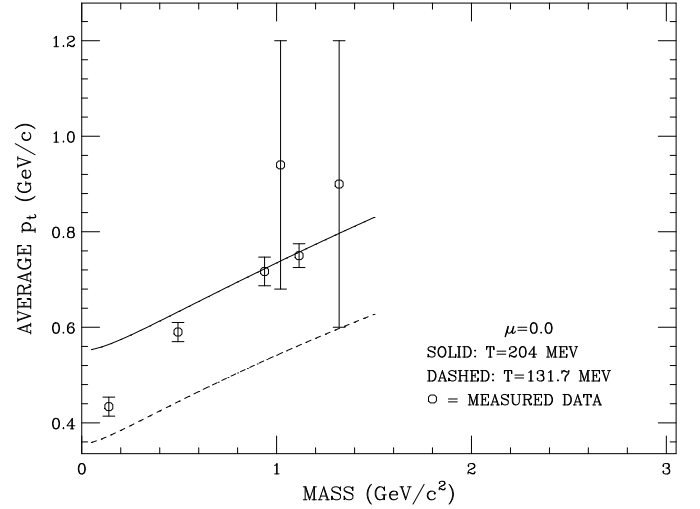}
\caption{\label{ptavg_nomu} The curves represent average transverse
momentum expected from a Bose-Einstein distribution with chemical
potential $\mu=0$. The solid curve is for temperature $T=204$ MeV and
the dashed curve is for $T=131.7$ MeV.}
\end{figure}

  An initial calculation was done using chemical potentials $\mu = 0$. Curves
of $<\!\!p_t\!\!>$ versus mass are plotted in Fig.~\ref{ptavg_nomu} for 
T=204 MeV and T=131.7 MeV, the latter being slightly more appropriate for
$\mu=0$ as in the fit of Fig.~\ref{zero_mu}.

 In Fig.~\ref{ptavg_nomu} it appears at first sight that the value of 
T=204 MeV has some amount of promise for being
useful and that T=131.7 MeV 
is completely inappropriate. Neither curve 
recognizes the more pronounced decrease 
with mass in the
experimental values of
$<\!\! p_t \!\!>$ for the kaon and pion.

\begin{figure}[h]
\includegraphics[height=3.5in,angle=90]{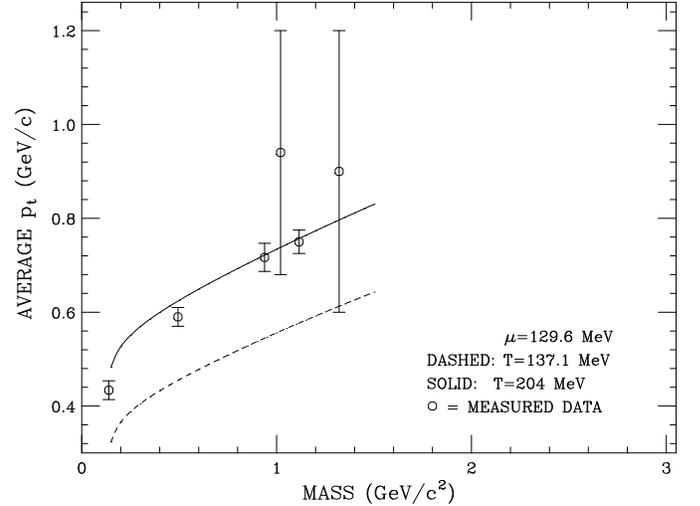}
\caption{\label{ptavg_mu} The curves represent average transverse
momentum expected from a Bose-Einstein distribution with chemical
potential $\mu=129.6$ MeV. The solid curve is for temperature $T=204$ MeV and
the dashed curve is for $T=137.1$ MeV.}
\end{figure}

This last feature is improved by using the chemical potential $\mu=129.6$ MeV.
Fig.~\ref{ptavg_mu} compares use of T=204 MeV and T=137.1 MeV, the 
latter number being more
relevant for the non zero value of $\mu$ as in Fig.~\ref{quark_mu}.
The higher temperature curve now comes closer to the measured values for
the kaon and pion, but it is still well outside experimental error bars.
The lower temperature curve remains unacceptable because of a nearly
uniform displacement below the data points.

  This displacement can be understood if one uses a suggestion
made by Levai and M\"uller~\cite{b18} which said that part of the transverse
momentum may be due to the transverse flow of the gas as a whole.
In order to calculate the effects of transverse flow it is advantageous
to define a particle density 4-flow vector $dN^{\nu}$. Within a 
constant factor this is given by~\cite{degroot} 

\begin{equation}
dN^{\nu}(p)= d^3{\bf p} (\frac{p^{\nu}}{p^0}) \frac{1}
{[e^{(p^{\nu}U_{\nu}-\mu)/T} \mp 1]},
\label{dN_nu}
\end{equation}
where $(p^{\nu}/p^0)$ is the $\nu$ velocity component of particles in
the infinitesimal group $dN^{\nu}$. The particle distribution function is 
here written in terms of an invariant $p^{\nu}U_{\nu}$. In this form
it is sometimes called the J\"uttner distribution~\cite{juttner1,juttner2}.
$U_{\nu}$ is the constant 4-flow velocity of the gas: 
$U_{\nu}=(\gamma_f c\,\,, -\gamma_f {\bf v_f})$,
where ${\bf v_f}$ is the flow velocity and $\gamma_f=1/\sqrt{1-v_f^2/c^2}$.
If the invariant is evaluated in the collision 
center of mass system for the case of
a particle with momentum {\bf p} parallel to the flow velocity ${\bf v_f}$,

\begin{equation}
p^{\nu}U_{\nu}=\gamma_f (E- \beta_f \, pc)=E',
\label{invar}
\end{equation}
where $E'$ is the particle energy in the rest system of the flowing gas. 

 We obtain an expression which is needed for calculating 
cross sections by forming
the invariant product
 
\begin{equation}
dN^{\nu}U_{\nu}= (\frac{d^3{\bf p}}{p^0}) \frac{p^{\nu}U_{\nu}}
{[e^{(p^{\nu}U_{\nu}-\mu)/T} \mp 1]}.
\label{dNU}
\end{equation}

When the flow velocity is zero, $v_f=0$, the invariant product 
reduces to dn as used in Eq.(\ref{int_eqn}). If the flow velocity is not
zero and is parallel to the transverse momentum $p_t$, the invariant
product becomes a more general expression for the invariant dn.

\begin{equation}
dn=\frac{p_t \, E'\,dp_t\,dy}{[e^{(E'-\mu)/T} \mp 1]}
\label{dn_flow}
\end{equation}
$E'$ is the energy of a particle in the rest system of the flowing gas and is
calculated using Eq.(\ref{invar}).

The cross section data points are not effective in constraining values of the
flow velocity $\beta_f=v_f/c$. Because of this we have estimated a value of 
$\beta_f$ by fitting the 4 best measurements of $<\!\!p_t\!\!>$, those of the
$\pi$, $K$, $p$, and $\Lambda$. Nevertheless it is still necessary to select
a reasonably correct value of T to be used in Eq.(\ref{avg_pt}) for computing
$<\!\!p_t\!\!>$. We find a value of $\beta_f=0.27 \pm 0.02$.
The error is merely statistical and in no way reflects 
the systematic uncertainty caused
by secondaries from resonance decays. 

\begin{figure}[h]
\includegraphics[height=3.5in,angle=90]{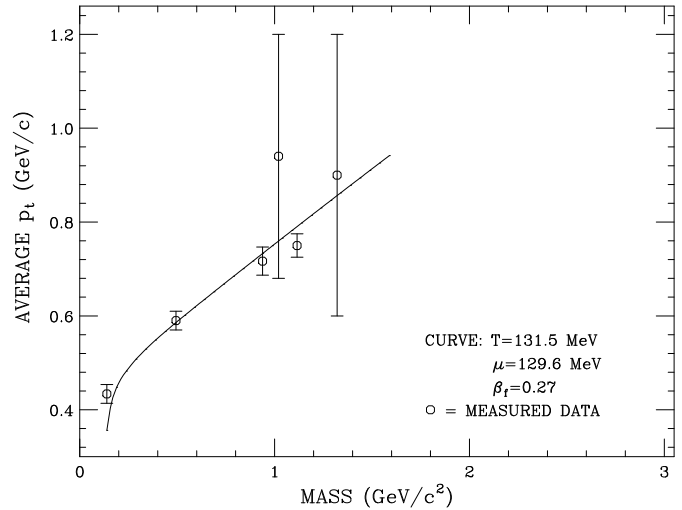}
\caption{\label{ptavg_flow} The solid curve represents average transverse
momentum expected from a J\"uttner distribution with $T=131.5$
MeV and chemical
potential $\mu=129.6$ MeV when momenta are boosted by a transverse
flow with $\beta_f=0.27$.}
\end{figure}

Some insight into the systematic error
might be gained by observing how close the fitted curve in Fig.~\ref{ptavg_flow}
comes to the $<\!\!p_t\!\!>$ measured for the proton, since the  
heavy mass of the proton causes recoil effects
in decays to be less likely to shift average values of momenta. 
The curve is well
within the statistical error on the measured value of $<\!\!p_t\!\!>$ for the proton.

  The curve in Fig.~\ref{ptavg_flow} was obtained using $T=131.5 \pm 3.4$ MeV. 
This is the temperature obtained if one fits the primary cross sections using
Eq.(\ref{dn_flow}) with $\beta_f=0.27$. 
Thus the cross section fit and the $\beta_f$
fit are consistent. For the cross section fit the 
normalizing constant (Eq.(\ref{sig_eqn}))
is $A=600 \pm 86$ mb/Ge$\mathrm{V}^3$ and the $\chi^2=3.57$. 
The fitted curve using these
values is imperceptibly
different from that shown in Fig.~\ref{quark_mu}.

\section{Transverse Momentum Spectra}

\begin{figure}[h]
\includegraphics[height=3.5in,angle=90]{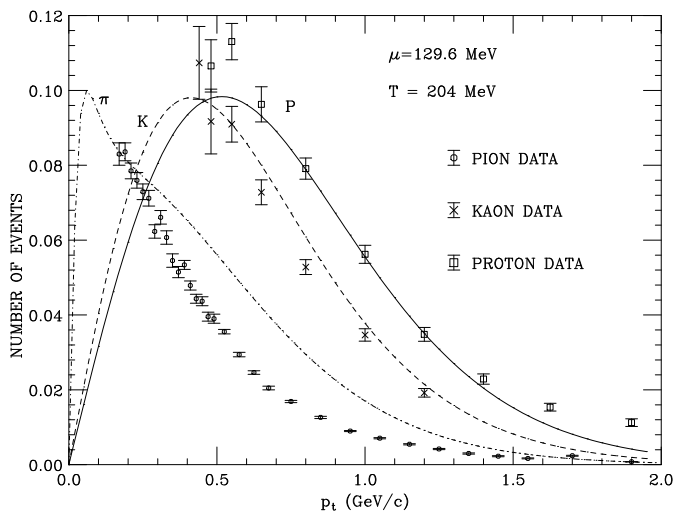}
\caption{\label{pdists_204} 
The curves represent transverse momentum distributions
for pions, kaons, and protons as expected from a Bose-Einstein distribution
with temperature $T=204$ MeV and chemical potential $\mu=129.6$ MeV.
Normalization was arbitrarily
chosen to fit all three curves within the plot. Data points  for the
3 particles were also normalized to facilitate comparisons with the 3 curves.}
\end{figure}

\begin{figure}[h]
\includegraphics[height=3.5in,angle=90]{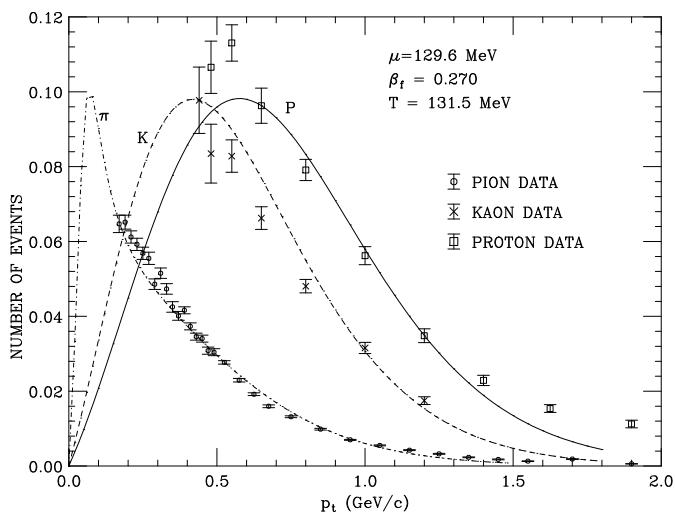}
\caption{\label{pdists_flow} The curves represent 
transverse momentum distributions
for pions, kaons, and protons as expected from a Bose-Einstein distribution
with temperature $T=131.5$ MeV, chemical potential $\mu=129.6$ MeV, 
and transverse flow  velocity $\beta_f = 0.27$. Normalization was arbitrarily
chosen to fit all three curves within the plot. Data points  for the
3 particles were also normalized to facilitate comparisons with the 3 curves.} 
\end{figure}

 Since the experimental $p_t$ spectra of the particles are a mixture
of primary particles and resonance decay products, it is a challenge to find a
useful means of comparing them to the statistical predictions for primary
particle production. The curves in Fig.~\ref{pdists_204} and
Fig.~\ref{pdists_flow} are statistical model predictions
for primary $p_t$ spectra of pions, kaons, and protons. Because
these particle cross sections vary by an order of magnitude, the vertical
scales for the curves have all been renormalized arbitrarily to
fit in the same plot. Experimental data points have also been renormalized
in order to help in making comparisons of data with model predictions.

  Figure~\ref{pdists_204} compares data points with a model that has T=204 MeV
and a chemical potential $\mu=129.6$ MeV. Data points 
are cut off sharply at low
momentum by absorption in the spectrometer material. 
The kaon and proton data are
not extremely different from the predicted primary curves, 
but the pion data points are
noticeably more condensed at low $p_t$ than this model predicts for the primary
pions.

Figure~\ref{pdists_flow} compares data points with a model 
that has T=131.5 MeV,
a chemical potential $\mu=129.6$ MeV, and a transverse flow velocity 
$\beta_f=0.27$.
All three particle spectra are similar to model predictions 
in an encouraging way.

\begin{figure}[h]
\includegraphics[height=3.5in,angle=90]{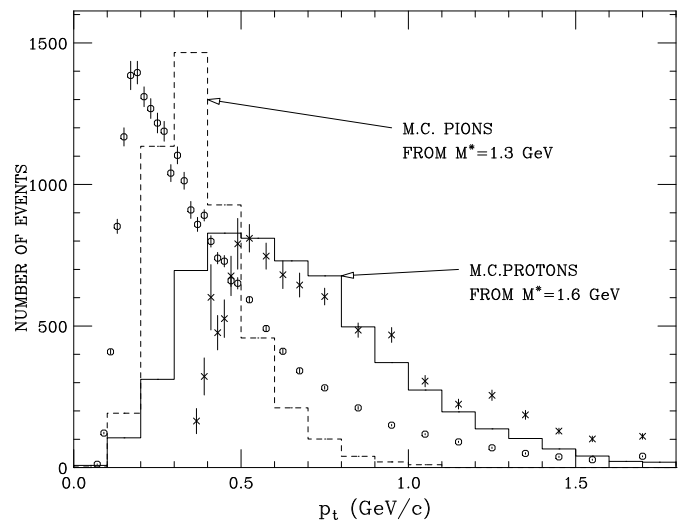}
\caption{\label{pdists_MC} Histograms are Monte Carlo generated distributions
using a gas with $T=204$ Mev. Dashed histogram represents pions from decay
of $M^*=1.3$ GeV to $\pi$p. Solid histogram represents protons from decay of
$M^*=1.6$ GeV to $\pi$p. Overall normalization is arbitrary. Data points
are normalized to give areas comparable to the histograms for comparison
purposes.} 
\end{figure}

  The influence of resonance decay products on the experimental 
$p_t$ spectra is perhaps
more complex than one might initially guess, especially for the spectrum of
the much lighter pions. We have examined spectra from the production and
decay of two ``typical'' resonances in our spectrometer using a Monte Carlo
program. The program generates particles at T=204 MeV using no chemical
potential and assumes a flat rapidity distribution. Experimental 
acceptance cuts and magnetic field are included in the analysis, 
but absorption effects are ignored.   
For pion sources the typical resonance is chosen to have a mass
$M^*=1.3$ GeV and a decay mode $M^* \rightarrow p+\pi$. For proton sources
the typical resonance is assumed to have a mass of  $M^*=1.6$ GeV and a 
decay mode of $M^* \rightarrow p+\pi$.  
The histograms in Fig.~\ref{pdists_MC} show the Monte Carlo spectra.
The {\it average} $p_t$ values for the 
Monte Carlo proton and pion spectra are quite close to the experimentally
measured values. 

  The shape of the pion spectrum shows a surprising peak 
at low momentum. Much of this peak comes from $M^*$ resonances initially
directed into the hemisphere opposite our detector. Backward pions from the
decay of this resonance pass through the aperture of the detector and are
processed as very low momentum pions. A small bump seen at $p_t \approx 250$
MeV/c in the experimental pion data might be associated with such an effect.

\section{Luminosity and Trigger Issues}

There are two methods used to calculate the luminosity of a
bunch crossing. One is the ``machine method'' and the other
is the ``cross section method''. The machine method consists
of first measuring and estimating properties of the particle
bunches and the particle accelerator and then using these to
calculate luminosity from first principles~\cite{b9}.
The cross section method counts the number of events from a 
beam crossing that result from a particular process and uses
the known cross section for that process in Eq.(\ref{lum})
in order to calculate the luminosity. Once the luminosity
is determined, all other cross section measurements are 
reduced to event counting experiments.

  When we first began to evaluate cross sections for the 
1988-1989 Tevatron run, the most accessible luminosities
were machine calculated values for which the data were posted at approximately
two hour intervals by the accelerator operators. We have used
these same machine luminosities~\cite{b9} to calculate all cross sections
presented in this paper. 

The trigger used to collect the data which we present from this
experiment was designed to select non-diffractive (nd) events.
Therefore a cross section computed for this trigger, 
$\sigma_{trig}$, should closely
approximate the best known value for the non-diffractive cross
section, $\sigma_{nd}$.

  Our primary trigger (PT) was typical of those used with collider detectors.
It required a triple coincidence between far upstream 
and far downstream counters
(BB) and the bunch crossing time T0. This coincidence was supplemented by
time based vetoes to eliminate beam-gas events from halo and satellite bunches.
Events selected by the PT trigger are called non-single-diffractive (nsd), 
because
elastic (el) and single-diffractive (sd) events are effectively elimnated by it.
Since our highest luminosity was around $2 \times 10^{28} \, 
\mathrm{cm}^{-2}\mathrm{sec}^{-1}$
(about two orders of magnitude below that of CDF), an interaction occured in 
about every 200 bunch crossings, and there was little need to consider cases 
where two sd events in the same bunch crossing could fake one nsd event.

  The trigger (S1) used for collecting data reported in this experiment made
use of the PT trigger described above in coincidence with a ``track'' in
the spectrometer arm. A track was defined by requiring drift chamber hits
in at least 3 of 4 possible planes both before and after the analysis magnet. 
This requirement of a spectrometer track near zero rapidity effectively
eliminated double-diffractive (dd) events from the event trigger (S1).

\begin{table}[h]
\caption{\label{sigmas}
This table contains cross sections used to check our
machine-derived luminosity against the more commonly used
cross-section-derived luminosities. Column 3 gives sources
for the cross sections listed in column 2.} 
\begin{ruledtabular}
\begin{tabular}{ccc}
Cross Section       & Value (mb)         &    Source  \\
\hline
$ \sigma_{trig}$    &  $41 \pm 6$        &    \cite{b6} \\
$ \sigma_{dd}$      &  $4.43 \pm 1.18$   &    \cite{b29} \\
$ \sigma_{sd}$      &  $9.46 \pm 0.44$   &    \cite{b30} \\
$ \sigma_{inel}$    &  $59.3 \pm 2.3 $   &    \cite{b31} \\
$ \sigma_{nd}$      &  $45.4 \pm 2.6 $   & \cite{b29},\cite{b30},\cite{b31} \\ 
\end{tabular}
\end{ruledtabular}
\end{table}

  The first row of Table~\ref{sigmas} gives the cross section of the S1 trigger
as computed using machine-determined luminosity for the C0 intersection,
$ \sigma_{trig}=41 \pm 6$ mb. We should compare $\sigma_{trig}$ to the
non-diffractive cross section $\sigma_{nd}$ as derived from 
cross-section-determined
luminosities given in Table~\ref{sigmas}.
\begin{equation}
\sigma_{nd}=\sigma_{inel}-\sigma_{dd}-\sigma_{sd}=45.4 \pm 2.6 \, \mathrm{mb}.
\label{sigs}
\end{equation} 
In this calculation we  have used an average value for $\sigma_{inel}$ which 
was adopted by the CDF and D0 collaborations for defining 
cross-section-determined luminosities~\cite{b31}.
Our machine-determined $\sigma_{trig}$ is about 10\% lower than the 
cross-section-determined $\sigma_{nd}$, 
but the two values agree within the errors.
Should one ever choose to compare cross sections in this paper with those
published by CDF or D0, then it might be desirable to shift one set or
the other systematically by 10\%.

It may be interesting to note that increasing all of our 
absolute cross sections
by 10\% would make the scale factor for our cross section fit 
agree almost exactly
with that for $\pi p$ scattering in Eq.(\ref{fermi}).

\section{Energy Dependence}

 Much can be learned about the properties of the thermal state by considering
the production properties of the $\phi$ meson. 
As one can see from Fig.~\ref{tableau_tst}(d),
there are negligible contributions to the inclusive $\phi$ cross section by
secondaries from higher mass decays. The $\phi$ cross section at y=0 {\it is}
essentially a primary cross section. If the same thermal production mechanism
prevails in other experiments, then a measurement of the phi cross section
in those experiments will allow a comparison of scales among experiments.
The decay of the $\phi$ into lepton pairs
and kaon pairs means that most experiments are equipped with a simple trigger
which can record its presence.

\begin{figure}[h]
\includegraphics[height=3.5in,angle=90]{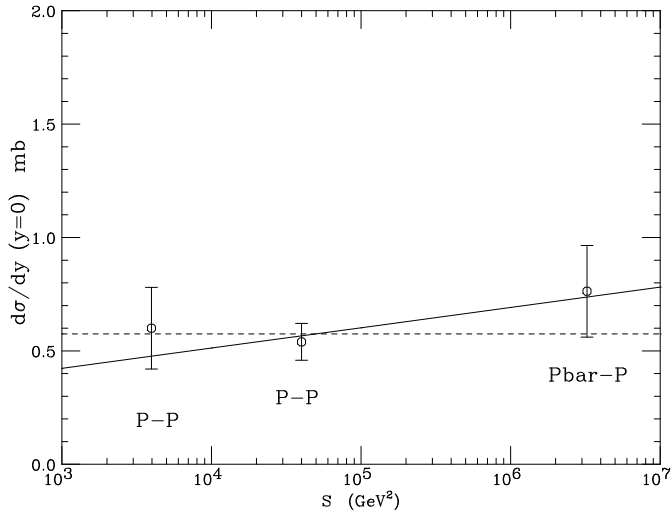}
\caption{\label{sig_log} Cross section 
$d\sigma/dy$ at y=0 for inclusive $\phi$ meson
production is plotted vs. s, the square of collision energy. 
Cross sections show
very little dependence on either energy or an annihilation option.}
\end{figure}

  Absolute cross sections at collider energies are rarely reported, but we are
able to plot two additional values of $d\sigma/dy|_{y=0}$ for 
$\phi$ mesons in Fig.~\ref{sig_log}
as a function of s~\cite{b32} . The lower energy cross sections 
are for pp collisions,
and the highest energy one is  our value at 
$\sqrt{s}=1800$ GeV for $\bar{p}p$ collisions.

  We have observed evidence for the thermal state only in collisions
with the potential for annihilation ($\bar{p}p$ and $e^+e^-$ initial states).
The similarity of the $pp$ and $\bar{p}p$ cross sections for $\phi$ mesons 
shown in Fig.~\ref{sig_log} suggests that 
the origin of the thermal state may be in
the interaction of gluons, which are common components 
of both protons and antiprotons.

  An inelastic cross section such as the total cross section for 
$\phi$ production 
might be expected to have an energy dependence which is bounded by some
relation like the Froissart limit~\cite{b33}, 
$\sigma_{tot}(\phi) \propto \ln^2s$. 
To the extent that inclusive cross sections have a flat rapidity distribution
for $|y|< y_{max}$, we can write 
$$\sigma_{tot}(\phi) \simeq 2\, y_{max} \frac{d\sigma}{dy}\Big|_{y=0} 
\simeq \ln (s/m^2) \,\frac{d\sigma}{dy}\Big|_{y=0}.$$ 
In this case it seems likely 
the $\phi$ cross section at y=0 would respect a reduced Froissart limit of 
$d\sigma/dy|_{y=0} \leq \ln s$.

  In Fig.~\ref{sig_log} we have plotted best fits to 
$$\frac{d\sigma}{dy}\Big|_{y=0}=\mathrm{constant} \qquad 
(=0.58 \pm 0.07\,\,\mathrm{mb})$$ and
 $$\frac{d \sigma}{dy}\Big|_{y=0}=A\, \ln(s/s_0) \quad (=0.039 \pm 0.005)
\cdot (\ln(s/m^2_{\pi})\,\, \mathrm{mb}),$$ 
where $s_0$ is an energy scale factor
to which  fits are insensitive. 
We have chosen $s_0$ to be the square of the pion mass.

  The $\ln s$ fit has a $\chi^2=0.59$, and the constant cross section fit 
has a $\chi^2=1.1$.
Both possibilities are acceptable for 2 degrees of freedom. 
The data are consistent with a
constant $\phi$ cross section or a slight increase with energy 
near a Froissart bound. 

\section{Nuclear Cross Sections}

  Although one can see in Fig.~\ref{kt_expfit} that 
the two nuclear cross sections
do not fit the thermal model well, they seem to exhibit a tendency to be in the
neighborhood of the dashed curve that describes the model. The cross section 
per spin state for
deuterium is a factor of 5.0 below the dashed curve, and the cross section 
per spin state for
tritium is a factor of 8.7 above the dashed curve. It should be mentioned that
these plotted points are estimates of the cross sections 
and not direct measurements.
We should ask if the estimate for the deuterium cross section could be
larger or if there might be a reason for the tritium cross section to
appear so large.

  The invariant cross sections, $(1/\pi)d^2\sigma/dy dp_t^2$, 
for $d^+$, $d^-$, and $t^+$
were measured directly in a very limited range of $p_t$, but values for the
integrated cross section $d\sigma/dy|_{y=0}$ had to be obtained without benefit
of fully measured distributions of
transverse momentum. The published estimates for deuterium and tritium
were made using variations of a Boltzman momentum distribution
which had adequately fit $p_t$ spectra for $K^-$ and $\bar p$ produced
at $\sqrt{s}=1.8$ TeV. Systematic errors were assigned based on the 
results of these variations. 

  The tritium cross section was based on 8 events satisfying experimental cuts.
Only one possible anti tritium event was observed. Anti tritium is more 
heavily absorbed by the spectrometer material, 
but the recording of events was heavily
prescaled by total multiplicity, sometimes as much as a factor of 26,
so at this level of statistics it is only suspicious that there
are not more anti tritium event candidates observed. 
 
 There are two reasons to consider why the tritium cross section 
plotted in Fig.~\ref{kt_expfit}
might be higher than expected. The first reason is that 
the observed tritium nuclei 
might not originate from the $p \bar p$ collisions. 
They might come instead from secondary
interactions in the wall of the 2 mm thick aluminum beam pipe.

  Approximately $2 \times 10^7$ particles entered 
the spectrometer arm during the
experiment. Some of these in principle could have ejected a tritium nucleus
from aluminum in a quasi elastic process such as
\begin{equation}
  \pi^+ +\, ^{27}Al \rightarrow \pi^+ +\, ^3H +\, ^{24}Mg\, ,
\label{24mg_eqn}
\end{equation}
\newline
where $^{24}Mg$ is a spectator nuclear fragment. Using only this process, 
we find $\sim 12\%$ of the spectrometer
particles have momentum high enough to produce the $^3H$ candidates we observe. 

   Spallation experiments have reported total cross sections of 27 mb 
for producing $^{24}Mg$  using
protons with momentum of 1.2 GeV/c~\cite{spallation}. 
In no  case have we found a published
rate for $^3H$ production in a spallation experiment, 
although some might have had the 
ability to observe it. Much of the production of light masses 
is expected to come from a
low energy evaporation process after the initial impact. 
However, if {\it all} of the 27 mb
cross section for producing $^{24}Mg$ proceeds by Eq.(\ref{24mg_eqn}) 
with $^3H$ in the forward
direction, then there would be $~ 800$ tritium candidates in the spectrometer. 
If $^3H$ fragments
exist in such numbers, it
cannot be safely assumed that all of these would be eliminated 
by the vertex cuts we used.

   There is a second reason the tritium point in Fig.~\ref{kt_expfit} 
might appear too high.
It could be that there are more contributing states to 
the cross section than the $(2J+1)$
spin states we assumed. The estimated cross section was divided 
by only $(2J+1)$ spin states
before plotting. 

   Some evidence exists that well defined ``nuclear'' states 
exist in which one of the
usual nucleons is replaced by a particle resonance. 
The best established examples of this
can be found in the $\Lambda^0$ hyperfragments, 
where the $\Lambda^0$ lives long enough to give
a distinctive decay signature in a detector. A collection 
of old data in reference~\cite{hyper_1}
lists 12 hypernuclei ranging from $^3H_{\Lambda}$ to $^{13}C_{\Lambda}$ 
and tabulates their
binding energies along with some observed decay modes. 
More recentlly high resolution
spectrometers~\cite{hyper_2} have resolved excited energy levels 
for the bound $\Lambda^0$ in
heavier nuclei. For $^{89}Y_{\Lambda}$ energy levels for s, p, d, and f 
$\,\Lambda^0$ orbitals
have been resolved.

   If $N^*$ resonances can also be bound in nuclei, 
there might be many modes which would contribute
to the observation of $^3H$ production in $p \bar p$ collisions. 
Although there
seem to be no experiments designed to explore this notion, 
one can find some indication 
of it. Inelastic cross sections for pions on tritium peak at a pion energy 
around 140 MeV~\cite{pi_3h},
suggesting the formation of $\Delta (1232)$, 
and final state particle ratios in $\pi^{\pm}$ 
collisions with $^3H$ and $^3He$  are consistent with dominance 
by the $\Delta$ baryon 
resonance~\cite{n_ratios}.

\section{Little Bang Nucleosynthesis}

  The Little Bang collision as observed in this experiment
takes place in approximately $10^{-26}$ seconds as the 
Lorentz contracted disks of the proton and antiproton pass through each other. 
Interfering pions continue to be emitted~\cite{b19} from the vicinity 
of the collision for
times $\sim 10^3$ longer than the collision time, 
since the measured size of the emission
volume has a ``radius'' of approximately 1 fm, 
somewhat larger than this in the longitudinal
direction than in the transverse direction. 
Slightly larger emission dimensions are
correlated with larger multiplicities of produced hadrons. 

  Several estimates~\cite{b19,rolf}
have been made of the energy density in the emission volume 
at the time of hadronization.
The values reported are sensitive both to the models used and the fraction
of observed final state particles assumed to originate from the emission volume.  
Energy
densities obtained range from $\sim 4$ to $\sim 60$ times that of a proton. 

  Primary cross sections per spin state for the light hadrons 
decrease with mass according to a Boltzmann distribution
having a characteristic temperture $T=204$ MeV. 
See Fig.~\ref{kt_expfit}. We can understand
this temperature in a qualitative way if we assume that the collision 
deposits energy into a volume of
hadronic size at the limit of the Uncertainty Principle. In that case
$\Delta E \, c\Delta t= \hbar c$, 
where $\Delta E$ is the energy deposited and $c\Delta t$
is a characteristic hadronic radius $r_h$.

\begin{equation}
 \Delta E = \hbar c/r_h
\label{h_uncert}
\end{equation}

   In the Uncertainty Principle $\Delta E$ is the measure of the 
drop-off energy of a probability
function. For a Boltzmann distribution 
we can identify $\Delta E$ with $T$~\cite{3_min}.

\begin{equation}
 T = \hbar c/r_h
\label{T_vs_rh}
\end{equation}
\newline 
If all produced hadrons share the same radius $r_h$, 
then $T$ is a constant which can be found experimentally.
A value of $T=204 \pm 14$ MeV implies 
a hadronic radius $r_h=0.97 \pm 0.07$ fm.
In some sense $r_h$ has to be an average radius. 

   In this experiment we observe the production of 
deuterium and anti deuterium
nuclei. The radius of a nucleus does not fit comfortably inside a hadronic 
radius, so its production must wait for the volume to expand to a larger radius
$r > r_h$. According to Eq.(\ref{T_vs_rh}) a larger radius will imply a 
correspondingly lower temperature and a smaller cross section 
than does the smaller radius
of a hadron having the
same mass as the nucleus. 

   Assuming a nuclear volume increases as the number of constituent 
hadrons, A, we can 
write an effective nuclear radius $r_A$ as

\begin{equation}
 r_A= r_h \, A^{1/3}.
\label{rN_vs_mN}
\end{equation}
\newline
The temperature for
nuclear production becomes $T_A=\hbar c/r_A$, and the cross section 
in millibarns per spin state 
for producing a nucleus $A$ becomes

\begin{equation}
 \frac{d \sigma(A)}{dy}\Big|_{y=0}= 45.3 \, e^{-m_A/T_A}.
\label{nucl_sig}
\end{equation}
\newline
Normalization at 45.3 mb fixes the primary proton cross section at the
value previously used with $T=204$ MeV. 

\begin{figure}[h]
\includegraphics[height=3.5in,angle=90]{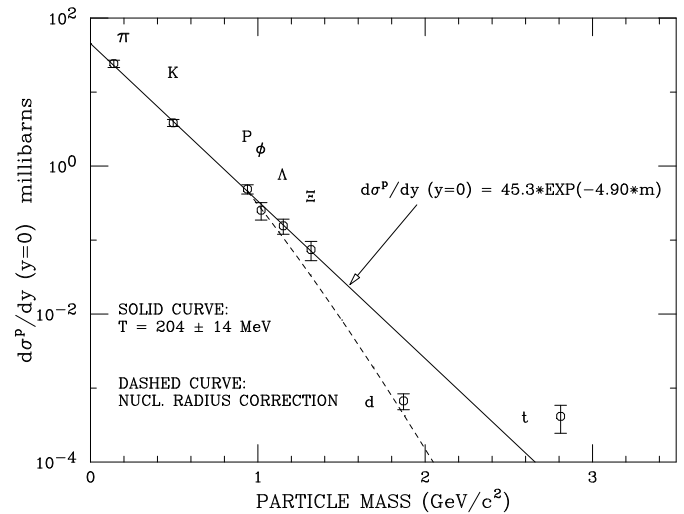}
\caption{\label{nucl_temp} Primary cross sections per spin state
are compared to a Boltzmann distribution in mass for a temperature
of $T=204$ Mev (solid curve) and to a distribution modified to
accomodate the radius of each individual nucleus (dashed curve).}  
\end{figure}

   A dashed curve in Fig.~\ref{nucl_temp}
plots the locus of nuclear cross sections per contributing state 
as predicted by Eq.(\ref{nucl_sig}).
The experimental estimates for the $d$ and $\bar d$ cross sections 
agree fairly well
with the prediction of a modified temperature for nuclear production, whereas
the experimental estimate of the $^3H$ cross section 
becomes $4 \times 10^3$ times
larger than the dashed curve.

   Nuclear cross sections decrease so rapidly with mass 
according to Eq.(\ref{nucl_sig})
that only the lightest nuclei might be detected at colliders. 
Table~\ref{nucl_table} lists predicted production rates relative to $\bar p$ 
rates for several light elements. 
Based on our
observation of 19 anti deuteron events at 
$\int{\cal L}dt=5.52 \pm 0.81 \, nb^{-1}$, we
estimate that at $\int{\cal L}dt=2100 \, pb^{-1}$ the CDF 
detector at the Tevatron 
could observe $\sim 3.6 \times 10^9$  anti deuteron events and 
$\sim 27$ anti $^4He$
events, provided there is no prescaling of the event triggers such as was
used in our experiment. 
This estimate allows for the increased geometrical acceptance
of the CDF detector but not for other detector specific features.

\begin{table}[h]
\caption{\label{nucl_table}
The production rate of a light nucleus relative to the production
rate of antiprotons is calculated using the temperature model of
Eq.(\ref{nucl_sig}). $J_A$ is the nuclear spin, and A is the 
mass number of the nucleus. Spin degeneracies are included in 
the calculation of the rate ratio R.}

\begin{ruledtabular}
\begin{tabular}{cccc}
Nucleus  &  A   & $J_A$& $R=N_A/N_p$ \\
\hline
 $^1H$       &   1   &    1/2     &    1.0  \\
 $^2H$       &   2   &     1      &  $1.38 \times 10^{-3}$  \\
 $^3H,\,^3He$&   3   &    1/2     &  $2.27 \times 10^{-7}$  \\
 $^4He$      &   4   &     0      &  $1.04 \times 10^{-11}$ \\
 $^6Li$      &   6   &     1      &  $2.52 \times 10^{-20}$ \\
 $^7Li$      &   7   &    3/2     &  $3.62 \times 10^{-25}$ \\
\end{tabular}
\end{ruledtabular}
\end{table}

  If nuclei are really formed in nucleon-nucleon collisions, 
it should be relatively
easy to verify the fact with suitable triggers 
in conjunction with one of the high resolution
detectors that currently exist.
It would be essential to require that the instrumentation not saturate on
pulse heights up to 10X-20X minimum ionization.
The high statistics expected for 
the lightest nuclei ($^2H$, $^3H$, $^3He$) would
permit one to  study the extent to which they are produced in pairs or to
put limits on the equality of nuclear and antinuclear cross sections.
Because a temperature of $T=204$ Mev also seems to organize the
hadron production cross sections from $e^+e^-$ collisions, it may be useful to
search for nuclear production in those data sets as well.

  Little Bang Nucleosynthesis (LBN) would seem to be quite different from 
Big Bang Nuleosynthesis (BBN)~\cite{Steigman}. BBN begins with a large
net baryon number, but LBN  begins with zero net baryon number, 
initiated perhaps by
annihilation or by gluon-gluon interaction. Electrons and photons participate
in BBN in a large way, but there is little evidence for either of these 
in LBN (see section IV).
In BBN nuclei are formed at temperatures 
comparable to nuclear binding energies. 
Nuclei from LBN appear to be formed somehow in an environment 
with temperatures  that
are many times larger than nuclear binding energies.

\begin{figure}[h]
\includegraphics[height=3.5in,angle=90]{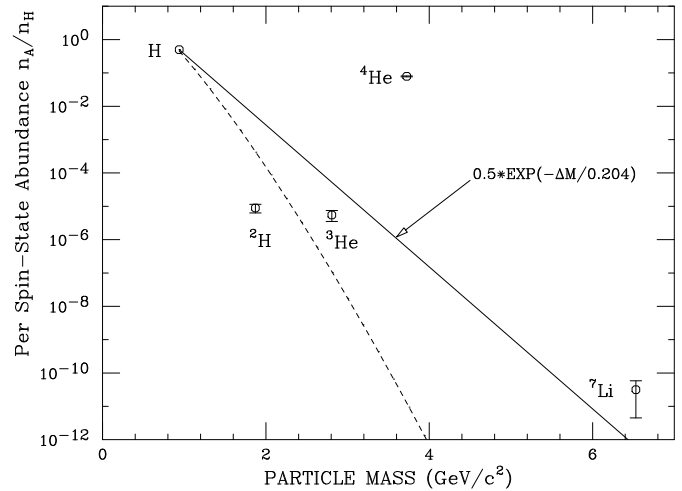}
\caption{\label{big_bang_sigs} Plotted points are relic cosmic
abundances of $^2H$, $^3He$, $^4He$, and $^7Li$ 
relative to  hydrogen~\cite{Steigman}
after dividing each ratio by $(2J+1)$. 
Variations of the Big Bang Nucleosynthesis
model strive to reproduce these data points. 
The solid curve when used with $\Delta M=m_A-m_p$ 
has the slope for $T=204$ MeV,
which fits elementary particle production in this paper. The dashed curve 
represents a
temperature modification which might accomodate nuclear size.}
\end{figure}

  Fig.~\ref{big_bang_sigs} has been included here to illustrate the extent to
which LBN fails to reproduce the relic cosmic abundances 
that BBN accounts for so
well~\cite{Steigman}. The ratio of cosmic abundances relative to
hydrogen have been divided by $(2J+1)$ for plotting. Nuclear curves from
Fig.~\ref{nucl_temp} have been inserted into the figure for comparison. 

\section{Conclusions}

   This paper has presented 9 absolute cross sections 
($d\sigma/dy|_{y=0}$) for particle
production at $90^0$ to colliding $p$ and $\bar p$ 
beams with $\sqrt{s}=1.8$ TeV. One
of these cross sections is for photon production. 
Two are for nuclear production. The
remaining six are for light elementary hadron production.

   The six hadron cross sections were corrected for 
secondary contributions from decays 
of higher masses in order to produce a set of six primary cross sections. 
When the primary 
cross sections are divided by the spin degeneracy $(2J+1)$ 
of individual particles, we
learn that the primary cross sections per spin state precisely 
follow an exponentially decreasing
distribution in particle rest mass with a temperature $T=204 \pm 14$ MeV. 
By summing over
all spin states, one finds the total primary cross section for a particle 
of specified charge, spin J, and rest
mass m is given by

\begin{equation}
\label{lambar_exp}
  \frac{d \sigma^p(\mathrm{tot})}{dy}\Big{|}_{y=0}=0.721 
\cdot(\pi \lambdabar_{\pi}^2)(2J+1) \cdot e^{-m/T}.
\end{equation}

Using an argument based on the Uncertainty Principle, 
we set $T=\hbar c/r_h$, where
$r_h$ is a hadronic radius given by $r_h=0.97 \pm 0.07$ fm. 
For nuclear production we
replace $r_h$ by $r_A=r_h \, A^{1/3}$ and find a temperature $T_A$. 
When $T_A$ is used in
Eq.(\ref{lambar_exp}), it yields an acceptable value for the 
anti deuteron cross section.  

This purely
thermal production mechanism exhibits no significant strangeness suppression,
implying that experimentally observed strangeness 
suppression is entirely explained by contributions 
from decay products of higher
masses. As discussed in section IV, we were unable to observe any
primary production at the photon and electron masses, 
so the thermal mechanism is probably
relevant  only for strongly interacting particles. 

  It appears there may be two temperatures that are useful in
describing the entire particle production
process. The temperature T=204 MeV precisely describes 
the primary particle ratios
in the initially produced state, 
and a temperature T=132 MeV describes the momentum
distributions in a final decay state in which pions have 
a uniquely low momentum.
An exothermic transition between the production state 
and decay state might account
for a transverse flow velocity of $\beta_f=0.27$. 
The average kinetic energy per
particle associated with this transverse flow is $\sim 20$ MeV, 
and thus it is not
a trivial amount.

   It is possible to observe the thermal production mechanism in $p \bar p$
collisions at $90^0$ (y=0)
because of its dominant cross sections in the low mass region.
For the production of masses in or above the charm sector
there is evidence that other mechanisms such as parton 
scattering become dominant.
The dominance may extend to many orders of magnitude at the highest particle
masses.

   We have examined cross sections for hadron production in the charm sector by
$e^+e^-$ collisions using published values of multiplicity fractions. 
To the degree
that decay branching ratios are well known, 
we find that primary $e^+e^-$ multiplicity
fractions also follow a Boltzmann distribution in mass for a temperature  
$T=204$ MeV.
There is no significant evidence for strangeness suppression 
in the
charm sector examined. 

   Since the $\phi$ meson is essentially a primary particle 
with negligible secondary
contributions, one can possibly gain some insight 
into the origin of the thermal 
mechanism by studying $\phi$ production. 
The $\phi$ cross section at $90^0$ is consistent with being the same
for $pp$ and $p \bar p$ collisions over a range of 
center of mass energies that vary
by a factor of $\sim 30$. This may suggest that gluon interactions rather
than annihilations are the essential source of the thermal production. 

\section{Acknowledgments}

 We would like to thank V. Barger, C. Goebel, T. Han, and S. Dasu for helpful
discussions. We are also indebted to Matt Herndon
for answering our numerous questions about the CDF
detector. O. Biebel made very helpful suggestions for data
sources. Useful hints on locating nuclear data were 
supplied by F. Ajzenberg-Selove and H. Taub.  

\section{Appendix}
\subsection*{Cross Section Estimates}

 We have used partial cross sections and raw data found in the
literature to make estimates of total inclusive cross sections for four
massive particles. The fact that experimenters themselves did not 
present total cross sections testifies to the difficulty and
large uncertainty associated with such calculations. Nevertheless
the estimates made below may be useful within an order of magnitude.

  $J/\psi$: Using a minimum bias trigger in a special run, CDF measured a
cross section~\cite{b22} for the prompt production of $J/\psi$ with a
value of
\begin{equation}
  \sigma (p_t>1.25,|y|<0.6)=2.86 \pm 0.01^{+0.34}_{-0.45} \, \mu b. 
\label{psi_sig}
\end{equation}

  Prompt $J/\psi$'s are those which are consistent with originating at the
beam line rather than from detached vertices. Since the entire $p_t$ spectrum
was measured for the $J/\psi$ in this experiment, 
we used that information to correct
the measured value upward by a factor of 1.28 before dividing by the effective
interval $\Delta y=1.2$. The result given below was divided by $(2J+1)=3$ and
plotted in Fig.~\ref{predict}.
\begin{equation}
 \frac{d\sigma(J/\psi)}{dy}\Big|_{y=0}=3.06 \,\, \mu\mathrm{b}. 
\label{psi_sigy}
\end{equation}

$D^{*+}$:  Using a detached vertex trigger, CDF measured the $D^{*+}$ 
cross section
as a function of $p_t$ above $p_t=6.0$ GeV/c~\cite{b23}.
We have used the $J/\psi$ momentum spectrum as a guide to increase this
cross section by a factor of 30.6 to account for the unmeasured
portion of the $p_t$ spectrum. Dividing by the effective rapidity interval
of $\Delta y=2$, we estimate
\begin{equation}
 \frac{d\sigma(D^{*+})}{dy}\Big|_{y=0}=0.075 \,\, \mathrm{mb}. \label{d*_sigy}
\end{equation}

This value is divided by $(2J+1)=3$ and plotted in Fig.~\ref{predict}.

$\Xi^-_b$: CDF has reported 17.5 of these events with a decay mode of
$\Xi^-_b \rightarrow  J/\psi\,\Xi^-$~\cite{b24}.

We divide this number by integrated luminosity $L=1.9 \,\mathrm{fb}^{-1}$ 
and multiply by
10 to correct for $J/\psi$ acceptance over the rapidity range when
$J/\psi$ momentum is $p_t > 3.0$ GeV/c. We multiply by 3.26 to correct
the unobserved $J/\psi$ momentum spectrum below $p_t=3.0$ GeV/c using the
$p_t$ spectra of reference \cite{b22} as a guide. We divide by 0.0593
to correct for the branching fraction of $J/\psi$ into the observed
$\mu^+\mu^-$ pair~\cite{b16} and multiply by $10^3$ to estimate a typical
probability for hadron decay into $J/\psi$. Dividing the resulting cross
section by $\Delta y=2$, we obtain
\begin{equation}
\frac{d\sigma(\Xi^-_b)}{dy}\Big|_{y=0}=2.53 \,\times\,10^{-6}\,\,\mathrm{mb}. 
\label{Xi_sigy}
\end{equation}

We have divided this number by $(2J+1)=2$ and plotted it in Fig.~\ref{predict}.

 $B^+$: CDF has measured the cross section at $\sqrt{s}=1.96$ TeV for
$p \bar p \rightarrow H_bX$, where X is anything and $H_b$ includes b hadrons
and anti b hadrons~\cite{b22}.
\begin{equation}
 \sigma(p\bar p \rightarrow H_bX,|y|<0.6)=17.6 \pm0.4^{+2.5}_{-2.3} 
\,\,\mu \mathrm{b}. \label{B_sig}
\end{equation}

We divide this by 2 to obtain a cross section only for hadrons and by
$\Delta y=1.2$ to estimate $d\sigma/dy$. The $B^+$ cross section can be
estimated using the branching fraction 
$B(b \rightarrow B^+)=0.398\pm0.012$~\cite{b16}
as a multiplier. The resulting cross section estimate for $B^+$ production is
\begin{equation}
\frac{d\sigma(B^+)}{dy}\Big|_{y=0}=2.92 \, \mu \mathrm{b}.  \label{B_sigy}
\end{equation}

We have further reduced this value by 10\% in an effort to scale it to 
a collision energy of $\sqrt{s}=1.8$ TeV
and divided by $(2J+1)=1$ for plotting in Fig.~\ref{predict}.


\begin{thebibliography}{99}
\bibitem{b36} Although the observed momentum spectrum for 
$B^+$ production is not 
exponential, it does not scale as the lower mass particles do in the momentum
range studied.
\end{thebibliography}
\end{document}